\pdfoutput = 1

\documentclass[a4paper,12pt]{article}
\usepackage{hyperref}

\usepackage{graphicx}
\usepackage{caption}
\usepackage{cite}
\usepackage{color}
\usepackage{subfig}
\usepackage{float}
\usepackage{slashed,dsfont}
\usepackage{longtable}
\usepackage{booktabs}
\usepackage{amsmath}
\usepackage{amssymb}

\usepackage{tabularx}
\usepackage{rotating}
\usepackage{lscape} 
\usepackage{epsfig}
\usepackage{color}
\usepackage{array}

\restylefloat{figure}

\usepackage[english]{babel}

\usepackage{geometry}
\numberwithin{equation}{section}
\usepackage{color}
\usepackage{array}
\newcolumntype{C}[1]{>{\centering\arraybackslash}p{#1}}
\usepackage{slashed,dsfont}
\usepackage{bm,wasysym}
\usepackage{longtable}
\usepackage{booktabs}

\usepackage{tabularx}
\usepackage{rotating}
\usepackage{lscape} 

\usepackage[normalem]{ulem}

\usepackage{graphicx}		
\usepackage{xcolor}

\usepackage{xspace}
\usepackage{ifthen}  
\usepackage{amssymb}
\usepackage{amsfonts}
\usepackage{upgreek} 
\usepackage{bigints}
\usepackage{lineno}  

\usepackage{hyperref}
\usepackage{cancel}
\RequirePackage[sort&compress,square,comma,numbers]{natbib}
\newboolean{uprightparticles}
\setboolean{uprightparticles}{false}

\def\ux85 {\mbox{UX85}\xspace}

\ifthenelse{\boolean{uprightparticles}}%
{

 \def\PDelta      {\ensuremath{\Delta}\xspace}                 
 \def\PXi      {\ensuremath{\Xi}\xspace}                 
 \def\PLambda      {\ensuremath{\Lambda}\xspace}                 
 \def\PSigma      {\ensuremath{\Sigma}\xspace}                 
 \def\POmega      {\ensuremath{\Omega}\xspace}                 
 \def\PUpsilon      {\ensuremath{\Upsilon}\xspace}                 
 
 \def\PB      {\ensuremath{\mathrm{B}}\xspace}                 
                  
 \def\PD      {\ensuremath{\mathrm{D}}\xspace}

 \def\PK      {\ensuremath{\mathrm{K}}\xspace}

 \def\Pi      {\ensuremath{\mathrm{i}}\xspace}

}
{

 \mathchardef\PDelta="7101
 \mathchardef\PXi="7104
 \mathchardef\PLambda="7103
 \mathchardef\PSigma="7106
 \mathchardef\POmega="710A
 \mathchardef\PUpsilon="7107
                  
 \def\PB      {\ensuremath{B}\xspace}                 
                  
 \def\PD      {\ensuremath{D}\xspace}

 \def\PK      {\ensuremath{K}\xspace}

 \def\Pi      {\ensuremath{i}\xspace}

}

\def\kaon  {\ensuremath{\PK}\xspace}
  \def\Kbar  {\kern 0.2em\overline{\kern -0.2em \PK}{}\xspace}

\def\Kz    {\ensuremath{\kaon^0}\xspace}
\def\Kzb   {\ensuremath{\Kbar^0}\xspace}
\def\KzKzb {\ensuremath{\Kz \kern -0.16em \Kzb}\xspace}
\def\Kp    {\ensuremath{\kaon^+}\xspace}
\def\Km    {\ensuremath{\kaon^-}\xspace}

\def\KpKm  {\ensuremath{\Kp \kern -0.16em \Km}\xspace}

  \def\Dbar    {\kern 0.2em\overline{\kern -0.2em \PD}{}\xspace}
\def\D       {\ensuremath{\PD}\xspace}

\def\Dz      {\ensuremath{\D^0}\xspace}
\def\Dzb     {\ensuremath{\Dbar^0}\xspace}
\def\DzDzb   {\ensuremath{\Dz {\kern -0.16em \Dzb}}\xspace}
\def\Dp      {\ensuremath{\D^+}\xspace}
\def\Dm      {\ensuremath{\D^-}\xspace}

\def\DpDm    {\ensuremath{\Dp {\kern -0.16em \Dm}}\xspace}

  \def\Bbar    {\kern 0.18em\overline{\kern -0.18em \PB}{}\xspace}

  \def\Y#1S{\ensuremath{\PUpsilon{(#1S)}}\xspace}

\def\Lz {\ensuremath{\PLambda}\xspace}
\def\Lbar {\ensuremath{\kern 0.1em\overline{\kern -0.1em\Lambda\kern -0.05em}\kern 0.05em{}}\xspace}

\def\Lb      {\ensuremath{\Lz_\bquark}\xspace}

\def\to                 {\ensuremath{\rightarrow}\xspace}

\newcommand{\mLb}{\ensuremath{m_{\Lb}}\xspace}

\def\qsq       {\ensuremath{q^2}\xspace}

\def\AT#1     {\ensuremath{A_{\mathrm{T}}^{#1}}\xspace}           

\def\C#1      {\ensuremath{\mathcal{C}_{#1}}\xspace}                       
\def\Cp#1     {\ensuremath{\mathcal{C}_{#1}^{'}}\xspace}                    
\def\Ceff#1   {\ensuremath{\mathcal{C}_{#1}^{\mathrm{(eff)}}}\xspace}        
\def\Cpeff#1  {\ensuremath{\mathcal{C}_{#1}^{'\mathrm{(eff)}}}\xspace}       
\def\Ope#1    {\ensuremath{\mathcal{O}_{#1}}\xspace}                       
\def\Opep#1   {\ensuremath{\mathcal{O}_{#1}^{'}}\xspace}                    

\newcommand{\bra}[1]{\ensuremath{\langle #1|}}             
\newcommand{\ket}[1]{\ensuremath{|#1\rangle}}              

\newcommand{\tev}{\ensuremath{\mathrm{\,Te\kern -0.1em V}}\xspace}
\newcommand{\gev}{\ensuremath{\mathrm{\,Ge\kern -0.1em V}}\xspace}
\newcommand{\mev}{\ensuremath{\mathrm{\,Me\kern -0.1em V}}\xspace}
\newcommand{\kev}{\ensuremath{\mathrm{\,ke\kern -0.1em V}}\xspace}
\newcommand{\ev}{\ensuremath{\mathrm{\,e\kern -0.1em V}}\xspace}
\newcommand{\gevc}{\ensuremath{{\mathrm{\,Ge\kern -0.1em V\!/}c}}\xspace}
\newcommand{\mevc}{\ensuremath{{\mathrm{\,Me\kern -0.1em V\!/}c}}\xspace}
\newcommand{\gevcc}{\ensuremath{{\mathrm{\,Ge\kern -0.1em V\!/}c^2}}\xspace}
\newcommand{\gevgevcccc}{\ensuremath{{\mathrm{\,Ge\kern -0.1em V^2\!/}c^4}}\xspace}
\newcommand{\mevcc}{\ensuremath{{\mathrm{\,Me\kern -0.1em V\!/}c^2}}\xspace}

\def\deriv {\ensuremath{\mathrm{d}}}

\def\gsim{{~\raise.15em\hbox{$>$}\kern-.85em
          \lower.35em\hbox{$\sim$}~}\xspace}
\def\lsim{{~\raise.15em\hbox{$<$}\kern-.85em
          \lower.35em\hbox{$\sim$}~}\xspace}

\def\tell1  {TELL1\xspace}
\def\ukl1   {UKL1\xspace}

\newboolean{articletitles}
\setboolean{articletitles}{true}
\allowdisplaybreaks[2]
\usepackage{mciteplus}

\newboolean{inbibliography}

\newcommand{\Lam}{\Lambda_{\rm QCD}}

\newcommand{\nn}{\nonumber}
\newcommand{\ord}{{\cal O}}

\def\Lb{{\Lambda_b}}
\def\Lam{{\Lambda}}
\def\mLb{{m_{\Lambda_b}}}
\def\mmLb{{m^2_{\Lambda_b}}}
\def\mL{{m_{\Lambda}}}
\def\mmL{{m^2_{\Lambda}}}

\def\plpl{{+\frac{1}{2}+\frac{1}{2}}}
\def\plmi{{+\frac{1}{2}-\frac{1}{2}}}
\def\mipl{{-\frac{1}{2}+\frac{1}{2}}}
\def\mimi{{-\frac{1}{2}-\frac{1}{2}}}
\def\re{{\rm Re}}  \def\im{{\rm Im}}
\def\mC{{\mathcal{C}}}
\def\mK{{\mathcal{K}}}

\definecolor{schrift}{RGB}{120,0,0}

\def\ARpe#1{{A^R_{\perp_{#1}}}}

\def\ARpe#1{{A^R_{\perp_{#1}}}}  \def\ARpa#1{{A^R_{\|_{#1}}}}  
        
\def\ARSPp{{A^R_{\rm S \perp}}} \def\ARSPm{{A^R_{\rm S \|}}}

\def\ALpe#1{{A^L_{\perp_{#1}}}}  \def\ALpa#1{{A^L_{\|_{#1}}}}

\def\ALSPp{{A^L_{\rm S \perp}}} \def\ALSPm{{A^L_{\rm S \|}}}

\def\AsRpe#1{{A^{\ast R}_{\perp_{#1}}}}  \def\AsRpa#1{{A^{\ast R}_{\|_{#1}}}}  
        
\def\AsRSPp{{A^{\ast R}_{\rm S \perp}}} \def\AsRSPm{{A^{\ast R}_{\rm S \|}}}

\def\AsLpe#1{{A^{\ast L}_{\perp_{#1}}}}  \def\AsLpa#1{{A^{\ast L}_{\|_{#1}}}}  
        
\def\AsLSPp{{A^{\ast L}_{\rm S \perp}}} \def\AsLSPm{{A^{\ast L}_{\rm S \|}}}

\begin{document}
\begin{titlepage}

\vspace{1.2cm}
\begin{center}
{\Large\bf\color{schrift} Polarized $\Lambda_b$ baryon decay to $p\pi$ and a dilepton pair}
\end{center}
		
\vspace{0.5cm}
\begin{center}
{\sc Diganta Das*, Ria Sain** } \\[0.3cm]
{\sf *Department of Physics and Astrophysics, University of Delhi, Delhi 110007, India}\\
{\sf **The Institute of Mathematical Sciences, Taramani, Chennai 600113, India}\\
{\sf **Homi Bhabha National Institute Training School Complex,}\\
 {Anushakti Nagar, Mumbai 400085, India}
\end{center}
		
\vspace{0.8cm}
\begin{abstract}
\vspace{0.2cm}\noindent
\noindent 
The recent anomalies in $b\to s\ell^+\ell^-$ transitions could originate from some New Physics beyond the Standard Model. Either to confirm or to rule out this assumption, more tests of $b\to s\ell^+\ell^-$ transition are needed. Polarized $\Lambda_b$ decay to a $\Lambda$ and a dilepton pair offers a plethora of observables that are suitable to discriminate New Physics from the Standard Model. In this paper, we present a full angular analysis of a polarized $\Lambda_b$ decay to a $\Lambda(\to p\pi)\ell^+\ell^-$ final state. The study is performed in a set of the operator where the Standard Model operator basis is supplemented with its chirality flipped counterparts, and new scalar and pseudoscalar operators. The full angular distribution is calculated by retaining the mass of the final state leptons. At the low hadronic recoil, we use the Heavy Quark Effective Theory framework to relate the hadronic form factors which lead to simplified expression of the angular observables where short- and long-distance physics factorize. Using the factorized expressions of the observables, we construct a number of test of short- and long-distance physics including null tests of the Standard Model and its chirality flipped counterparts that can be carried out using experimental data.

\end{abstract}
\end{titlepage}

\newpage
\pagenumbering{arabic}
\section{Introduction}
Several experimental results in the rare $b\to s\ell^+\ell^-$ processes have shown deviations from the Standard Model (SM) predictions. In the $B\to K^{(\ast)}\ell^+\ell^-$ decay the deviations in the $R_{K^{(\ast)}}$ observables hints to a possible violation of lepton flavor universality \cite{Aaij:2019wad, Aaij:2017vbb}. Moreover, the branching ratios of $B\to K\mu^+\mu^-$ \cite{Aaij:2014pli}, $B\to K^\ast\mu^+\mu^-$ \cite{Aaij:2013iag, Aaij:2016flj}, $B_s\to\phi\mu^+\mu^-$ \cite{Aaij:2015esa}, and the optimized observables in $B\to K^\ast\mu^+\mu^-$ decay \cite{Aaij:2020nrf} show systematic deviation from the SM predictions. Though inconclusive till now, New Physics (NP) beyond the SM could be the origin of these deviations.

LHCb is now capable to study $b\to s\ell^+\ell^-$ transitions in baryonic decays. Interestingly, the LHCb measurement of $\Lambda_b\to \Lambda\ell^+\ell^-$ branching ratio \cite{Aaij:2015xza} shows deviations from the SM expectations with the same trend as its mesonic counterparts. Phenomenologically, the $\Lambda_b\to\Lambda(\to p\pi)\mu^+\mu^-$ decay could be richer than its mesonic counterpart as the $\Lambda_b$ can be produced in polarized state. If the $\Lambda_b$ is polarized then the full angular distribution of $\Lambda_b\to \Lambda(\to p\pi)\ell^+\ell^-$ in the SM gives access to 34 angular observables \cite{Blake:2017une} which were recently measured by the LHCb \cite{Aaij:2018gwm}. In the LHCb, the polarization of $\Lambda_b$ was measured as $P_{\Lb}=0.06\pm 0.07\pm 0.02$ at $\sqrt{s}=7$ TeV \cite{Aaij:2013oxa}. At the CMS detector, $P_{\Lb}=0.00\pm 0.06\pm 0.02$ was measured at $\sqrt{s}=7$ and 8 TeV \cite{Aaij:2013oxa,CMS:2016iaf}. In future $e^+e^-$ collider, $\Lambda_b$ can be produced in longitudinal polarization also.

There are several theoretical studies of $\Lambda_b\to\Lambda\ell^+\ell^-$ in the SM. Using QCD sum rules to calculate the $\Lambda_b\to \Lambda$ transition, the unpolarized $\Lambda_b\to\Lambda\ell^+\ell^-$ was studied in reference \cite{Huang:1998ek}. In \cite{Feldmann:2011xf} a sum-rule analysis of the spectator-scattering corrections to the $\Lambda_b\to\Lambda$ form factors at large recoil were given. Further theoretical understanding of light-cone distribution amplitude of $\Lambda_b$ wave function was achieved in \cite{Ali:2012pn,Bell:2013tfa,Braun:2014npa}. Light-cone sum-rule calculations of the $\Lambda_b\to \Lambda$ form factors have been done in \cite{Wang:2008sm, Wang:2015ndk} and most recently, lattice QCD calculations of the same have been done in \cite{Detmold:2016pkz}. A covariant constituent quark model analysis of unpolarized $\Lambda_b\to\Lambda\ell^+\ell^-$ decay can be found in \cite{Gutsche:2013pp}. A full angular distribution of unpolarized $\Lambda_b\to\Lambda(\to p\pi)\ell^+\ell^-$ in the SM was first given in \cite{Boer:2014kda}. Model-independent new physics analysis for unpolarized $\Lambda_b\to\Lambda^{(\ast)}(\to N\pi)\ell^+\ell^-$ decay including a complete set of dimension-six operators and retaining the final state lepton masses were performed in \cite{Das:2018sms, Das:2018iap, Das:2020cpv, Roy:2017dum,Yan:2019tgn}. A full angular distribution of polarized $\Lambda_b\to\Lambda(\to p\pi)\ell^+\ell^+$ decay in the SM was first discussed in reference \cite{Blake:2017une}. In this paper we revisit the polarized $\Lambda_b\to \Lambda(\to p\pi)\ell^+\ell^-$ decay and extend the previous SM angular distribution by supplementing the SM operator basis by its chirality-flipped counterparts (henceforth SM$^\prime$), and new scalar and pseudoscalar operators (henceforth SP operators). We also retain the masses of the final state leptons which was neglected in the previous calculations \cite{Blake:2017une}. In the presence of the SP operators, we find two additional angular observables that are helicity suppressed. All the 36 angular observables depend on ten form factors that have been calculated at large recoil in the framework of light-cone-sum-rules \cite{Wang:2008sm, Mannel:2011xg, Wang:2015ndk}. The form factors also have been calculated in the lattice QCD \cite{Detmold:2016pkz} which is valid at low recoil. At low recoil, or large dilepton invariant mass squared $q^2$, the decay can be described by a Heavy Quark Effective Theory (HQET) framework \cite{Isgur:1989ed,Isgur:1990pm,Isgur:1989vq}, which ensures relations between the form factors known as Isgur-Wise relations \cite{Isgur:1990kf,Mannel:1990vg, Grinstein:2004vb}. The Isgur-Wise relations lead to the factorization of short- and long-distance physics in the angular observables. We use the factorized expressions  to construct several new clean tests of form factors and short-distance physics in the SM. We also discuss how these tests are affected by the presence of SM$^\prime$ and SP operators. Additionally, we construct for the first time combinations of observables that are sensitive to scalar NP only and therefore serve as null tests of the SM+SM$^\prime$. We also present a numerical analysis of several observables in the SM and NP scenarios.    

The organization of the paper is as follows: in Sec.~\ref{sec:effHam} we discuss the effective operators for $b\to s\ell^+\ell^-$ transition and discuss the derivations of the decay amplitudes. In Sec.~\ref{sec:angdist} we derived the full angular distribution of $\Lambda_b\to\Lambda(\to p\pi)\ell^+\ell^-$ for a polarized $\Lambda_b$ and discuss how to extract angular observables from the angular distribution. In section \ref{sec:HQET} we discuss the relations between form factors in HQET. The low-recoil simplifications of the angular observables due to the HQET and its consequences on the tests of short- and long-distance physics are discussed in section \ref{sec:lowrecoil}. A numerical analysis is presented in Sec.~\ref{sec:numerical} and our results are summarized in Sec.~\ref{sec:summary}. We also give several appendices where details of the derivations can be found.

\section{Effective Hamiltonian \label{sec:effHam}}
In the SM the rare $b \to s\ell^+\ell^-$ transition proceeds through loop diagrams which are described by radiative operator $\mathcal{O}_7$, semileptonic operators $\mathcal{O}_{9,10}$, and hadronic operators $\mathcal{O}_{1-6,8}$. The operators $\mathcal{O}_{7,9,10}$ are the dominant ones in the SM and read 
\begin{eqnarray}\label{eq:opbasisSM}
&O_7^{} = \frac{m_b}{e} \big[\bar{s}\sigma^{\mu\nu}P_{R}b\big]F_{\mu\nu}\, ,\quad \mathcal{O}_9 = \big[\bar{s}\gamma^\mu P_{L}b \big]\big[\bar{\ell} \gamma_\mu\ell \big] and 
\mathcal{O}_{10} = \big[\bar{s}\gamma^\mu P_{L}b \big]\big[\bar{\ell} \gamma_\mu\gamma_5\ell \big] .
\end{eqnarray}
The corresponding Wilson coefficients are $\mC_7$, $\mC_9$ and $\mC_{10}$. We assume only the factorizable quark loop corrections to the hadronic operators $\mathcal{O}_{1-6,8}$ which are absorbed into the Wilson coefficients $\mC_{7,9}$ (often written as $\mC_7^{\rm eff}$, $\mC_9^{\rm eff}$ in the literature). For simplicity, we ignore the non-factorizable corrections which are expected to play a significant role, particularly at large recoil or low dilepton invariant mass squared $q^2$ \cite{Beneke:2001at,Beneke:2004dp}.

Beyond the SM, effects of NP operators with the same Dirac structure as $\mathcal{O}_{9,10}$ can be trivially included by the modifications $\mC_{9,10}\to \mC_{9,10}+\delta \mC_{9,10}$. We additionally include the chirality flipped counterparts of $\mathcal{O}_{9,10}$ and new scalar and pseudoscalar operators 
\begin{eqnarray}\label{eq:opbasisNP}
\begin{split}
&\mathcal{O}_{9^\prime} = \big[\bar{s}\gamma^\mu P_{R}b \big]\big[\bar{\ell} \gamma_\mu\ell \big]\, ,\quad
\mathcal{O}_{10^\prime} = \big[\bar{s}\gamma^\mu P_{R}b \big]\big[\bar{\ell} \gamma_\mu\gamma_5\ell \big]\, \\
&\mathcal{O}_{S^{(\prime)}} = \big[\bar{s}P_{R(L)}b \big]\big[\bar{\ell} \ell \big]\, ,\quad
\mathcal{O}_{P^{(\prime)}} = \big[\bar{s}P_{R(L)}b \big]\big[\bar{\ell} \gamma_5\ell \big] .
\end{split}
\end{eqnarray}
The Wilson coefficients corresponding to $\mathcal{O}_{9^\prime,10^\prime}$, $\mathcal{O}_{S^{(\prime)},P^{(\prime)}}$ are $\mC_{9^\prime,10^\prime}$ and $\mC_{S^{(\prime)},P^{(\prime)}}$, respectively. NP operators of the tensorial structure have been ignored for simplicity as these operators lead to a large number of terms in the angular distributions that will be discussed elsewhere. The $\Lambda_b \to\Lambda$ hadronic matrix elements for the set of operators \eqref{eq:opbasisSM} and \eqref{eq:opbasisNP} are conveniently defined in terms of ten $q^2$ dependent helicity form factors $f^{V,A}_{0,\perp,t}$, $f^{T,T5}_{0,\perp}$ \cite{Feldmann:2011xf}.

Assuming factorization between the hadronic and leptonic parts, and neglecting contributions proportional to $V_{ub}V^\ast_{us}$, the amplitude of the decay process $\Lambda_b(p,s_p)\to \Lambda(k,s_k)$ $\ell^+(q_+)\ell^-(q_-)$ can be written as 
\begin{eqnarray}\label{eq:Mll}
	\mathcal{M}^{\lambda_1,\lambda_2}(s_p,s_k) &=& - \frac{G_F}{\sqrt{2}}V_{tb}V_{ts}^\ast \frac{\alpha_e}{4\pi} \sum_{i=L,R}\bigg[ \sum_{\lambda} \eta_\lambda H^{i,s_p,s_k}_{\rm VA, \lambda} L^{\lambda_1,\lambda_2}_{i,\lambda} + H^{i,s_p,s_k}_{\rm SP} L^{\lambda_1,\lambda_2}_i  \bigg]\, ,
\end{eqnarray}
where $p,k,q_+, q_-$ are the momentum of $\Lambda_b$, $\Lambda$, $\ell^+$ and $\ell^-$, respectively, and $s_p, s_k$ are the projection of $\Lambda_b$ and $\Lambda$ spins on to the $z$-axis in their respective rest frames. The polarization of the two leptons are denoted by $\lambda_{1,2}$, $\lambda=0,\pm 1, t$ are the polarization states of virtual gauge boson that decays to the dilepton pair, and $\eta_{\pm 1,0}=-1$, $\eta_t=+1$. The expressions of the hadronic matrix elements $H^{L(R)}_{\rm VA,\lambda}(s_\Lb,s_\Lam)$ and $H^{L(R)}_{\rm SP}(s_\Lb,s_\Lam)$ can be found in Ref.~\cite{Das:2018sms}. In the literature, the angular observables are usually expressed in terms of transversity amplitudes. For SM+SM$^\prime$ set of operators the transversity amplitudes are \cite{Das:2018iap} 
\begin{eqnarray}
A^{L(R)}_{\perp_1} &=& -\sqrt{2}N \bigg( f^V_\perp \sqrt{2s_-} \mC^{L(R)}_{\rm VA+} + \frac{2m_b}{q^2} f^T_\perp (\mLb + \mL) \sqrt{2s_-} \mC_7^{\rm eff} \bigg)\, ,\\
A^{L(R)}_{\|_1} &=& \sqrt{2}N \bigg( f^A_\perp \sqrt{2s_+} \mC^{L(R)}_{\rm VA-} + \frac{2m_b}{q^2} f^{T5}_\perp (\mLb - \mL) \sqrt{2s_+} \mC_7^{\rm eff} \bigg)\, ,\\
A^{L(R)}_{\perp_0} &=& \sqrt{2}N \bigg( f^V_0 (\mLb + \mL) \sqrt{\frac{s_-}{q^2}} \mC^{L(R)}_{\rm VA+} + \frac{2m_b}{q^2} f^T_0\sqrt{q^2s_-} \mC_7^{\rm eff} \bigg)\, ,\\
A^{L(R)}_{\|_0} &=& -\sqrt{2}N \bigg( f^A_0 (\mLb - \mL) \sqrt{\frac{s_+}{q^2}} \mC^{L(R)}_{\rm VA-} + \frac{2m_b}{q^2} f^{T5}_0\sqrt{q^2s_+} \mC_7^{\rm eff} \bigg)\, ,\\
A_{\perp t} &=& - 2\sqrt{2}N (\mC_{10} + \mC_{10^\prime})  (\mLb-\mL)\sqrt{\frac{s_+}{q^2}} f^V_t \, ,\\ 
A_{\| t} &=&  2\sqrt{2}N (\mC_{10} - \mC_{10^\prime})  (\mLb+\mL)\sqrt{\frac{s_-}{q^2}} f^A_t\, ,
\end{eqnarray}
where $s_\pm = (\mLb \pm \mL)^2 - q^2$, and the normalization constant is given by 
%
\begin{equation}
N(q^2) = G_F V_{tb}V_{ts}^\ast \alpha_e \sqrt{\tau_{\Lambda_b} \frac{q^2\sqrt{\lambda(\mmLb,\mmL,q^2)}}{3\cdot2^{11} m^3_{\Lambda_b} \pi^5 }\beta_\ell}\, ,\quad \beta_\ell = \sqrt{1 - \frac{4m_\ell^2}{q^2}}\, .
\end{equation}
The combination of Wilson coefficients $\mC^{L(R)}_{\rm VA\pm}$ are
\begin{align}
& \mC^{L(R)}_{\rm VA+} = (\mC_9 + \mC_{9^\prime}) \mp (\mC_{10} + \mC_{10^\prime})  \, ,\\
&\mC^{L(R)}_{\rm VA-} =  (\mC_9 - \mC_{9^\prime}) \mp (\mC_{10} - \mC_{10^\prime})   \, .
\end{align}
For the SP operators we follow the definition of transversity amplitudes from \cite{Das:2020cpv} and using the scalar helicity amplitudes given in \cite{Das:2018sms} we get
\begin{eqnarray}
\label{TA:scalar1}
A_{\perp \rm S}^{L(R)} &=& \sqrt{2}N f_t^V \frac{m_{\Lambda_b} - m_{\Lambda}}{m_b - m_s} \mC_{\rm SP +}^{L(R)}\, ,\\
\label{TA:scalar2}
A_{\| \rm S}^{L(R)} &=& -\sqrt{2}N f_t^A \frac{m_{\Lambda_b} + m_{\Lambda}}{m_b + m_s} \mC_{\rm SP -}^{L(R)}\, ,
\end{eqnarray}
where the Wilson coefficients are 
\begin{align}
&\mC_{\rm SP +}^{L(R)}= (\mC_{S} + \mC_{S^\prime}) \mp (\mC_P + \mC_{P^\prime})\, , \\
&\mC_{\rm SP -}^{L(R)}=(\mC_S - \mC_{S^\prime}) \mp (\mC_P - \mC_{P^\prime}) \, .
\end{align}
The subsequent parity violating weak decay amplitudes of $\Lambda(k,s_k)\to p(k_1)\pi(k_2)$ are calculated following Ref.~\cite{Boer:2014kda} and using the baryon spinor given in Appendix  \ref{sec:LamRF}. In the angular observables however, it is the parity violating parameter $\alpha_\Lambda=0.642\pm 0.013$ \cite{Patrignani:2016xqp} that is relevant.

The leptonic helicity amplitudes are defined as
\begin{eqnarray}
L^{\lambda_1\lambda_2}_{L(R)} &=& \langle\bar{\ell}(\lambda_1)\ell(\lambda_2)|\bar{\ell}(1\mp\gamma_5)\ell|0\rangle\, ,\\
L^{\lambda_1\lambda_2}_{L(R),\lambda} &=&\bar{\epsilon}^\mu(\lambda) \langle\bar{\ell}(\lambda_1)\ell(\lambda_2)|\bar{\ell}\gamma_\mu(1\mp\gamma_5)\ell|0\rangle\, ,
\end{eqnarray}
where $\epsilon^\mu$ is the polarization of the virtual gauge boson that decays to the dilepton pair. The detailed expressions of the amplitudes are given in Appendix \ref{sec:LepHel}.  

\begin{figure}[h!]
	\begin{center}
		\includegraphics[scale=0.8]{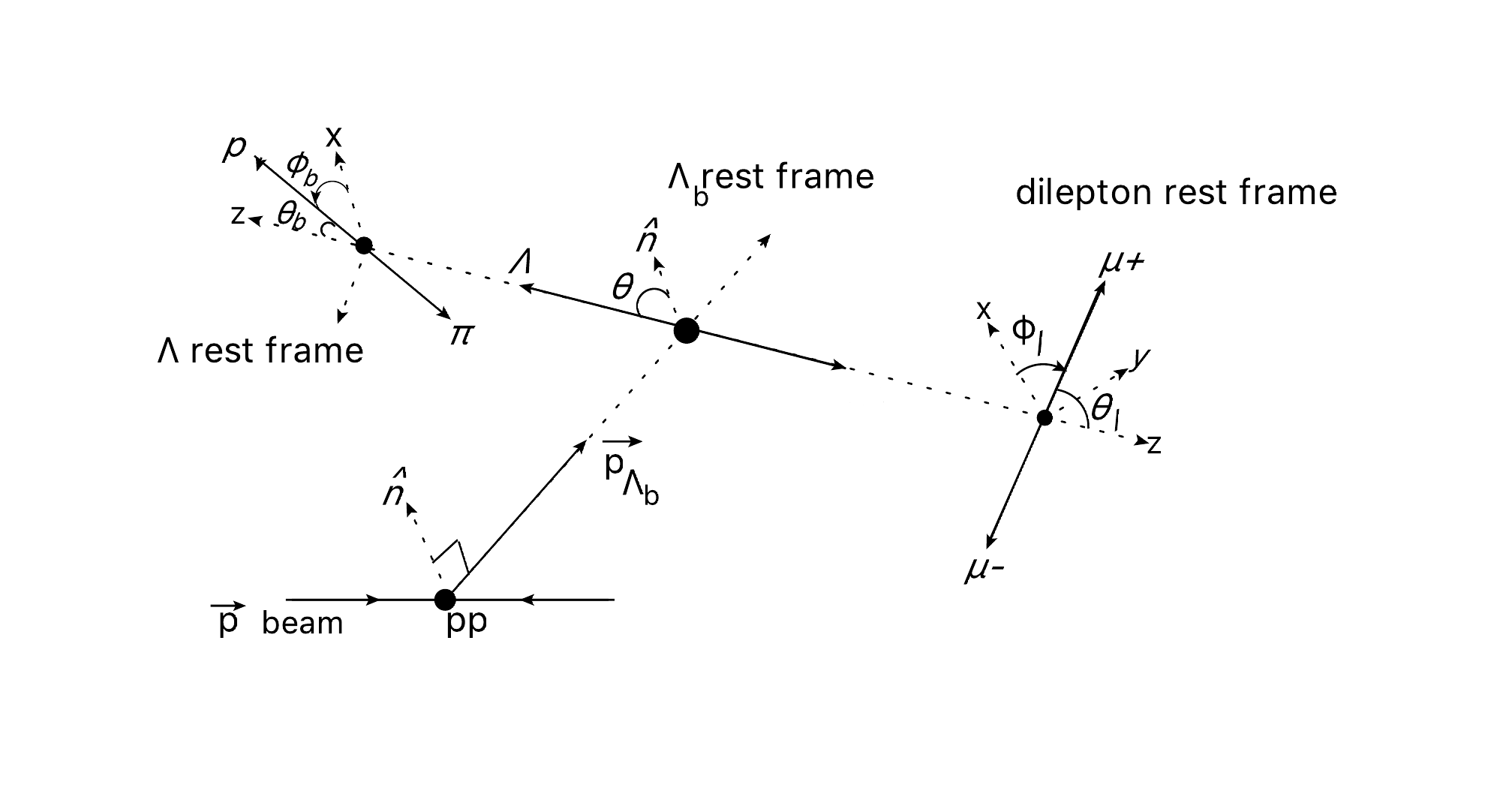}
		\caption{Definition of the angular basis for decay of polarized $\Lambda_b\to\Lambda(\to p\pi)\ell^+\ell^-$  \label{Fig:angular}}
	\end{center}
\end{figure}
\section{Angular Distributions\label{sec:angdist}} 
The set of angles that forms the angular basis are $\theta$, $\theta_{b,\ell}$ and $\phi_{b,\ell}$ \cite{Aaij:2018gwm}. Since transverse polarization is considered, it is appropriate to define a normal vector $\hat{n}=\hat{p}_{\rm beam}^{\rm \{lab\}} \times \hat{p}_{\Lambda_b}^{\rm \{lab\}}$ where $\hat{p}_{\rm beam}^{\rm \{lab\}}, \hat{p}_{\Lambda_b}^{\rm \{lab\}}$ are unit vectors in the lab frame. The angle $\theta$ between the $\hat{n}$ and the direction of $\Lambda$, in the $\Lambda_b$ rest frame is defined as $\cos\theta=\hat{n}.\hat{p}_\Lambda^{\{\Lambda_b \}}$. To describe the decay of $\Lambda\to p\pi$ and the dilepton system, we construct coordinate system $\{\hat{z}_b, \hat{y}_b, \hat{x}_b \}$ and $\{\hat{z}_\ell, \hat{y}_\ell, \hat{x}_\ell \}$, respectively. The $z$-axes are as follows: $\hat{z}_b=\hat{p}_\Lambda^{\{\Lambda_b\}}$, $\hat{z}_\ell=\hat{p}_{\ell^+\ell^-}^{\{\Lambda_b\}}$. The other two axes are defined as: $\hat{y}_{b,\ell} = \hat{n}\times \hat{z}_{b,\ell}$ and $\hat{x}_{b,\ell} = \hat{n}\times \hat{y}_{b,\ell}$. The angles $\theta_\ell$ and $\phi_\ell$ are made by the $\ell^+$ in the dilepton rest frame and the angles $\theta_b$  and $\phi_b$ are made by the proton in the $\Lambda$ rest frame as shown in figure \ref{Fig:angular}. With this angular basis, the six fold angular distribution of $\Lambda_b\to\Lambda(\to p\pi)\ell^+\ell^-$ for the ${\rm SM+SM}^{\prime}+{\rm SP}$ set of operators is
\begin{align}\label{eq:angular}
\frac{\deriv^{6}\mathcal{B}}{\deriv\qsq\,\deriv\vec{\Omega}(\theta_\ell,\phi_\ell,\theta_b,\phi_b,\theta)} = \frac{3}{32\pi^{2}} \Big(
& \left(K_1\sin^2\theta_\ell+K_2\cos^2\theta_\ell+K_3\cos\theta_\ell \right)  +  \nn\\[-5pt]
& \left(K_4\sin^2\theta_\ell+K_5\cos^2\theta_\ell+K_6\cos\theta_\ell\right)\cos\theta_b +  \nn\\
& \left(K_7\sin\theta_\ell\cos\theta_\ell+K_8\sin\theta_\ell\right)\sin\theta_b\cos\left(\phi_b+\phi_\ell\right) +  \nn\\
&\left(K_9\sin\theta_\ell\cos\theta_\ell+K_{10}\sin\theta_\ell\right)\sin\theta_b\sin\left(\phi_b+\phi_\ell\right) +  \nn\\
& \left(K_{11}\sin^2\theta_\ell+K_{12}\cos^2\theta_\ell+K_{13}\cos\theta_\ell\right)\cos\theta +  \nn\\
& \left( K_{14}\sin^2\theta_\ell+K_{15}\cos^2\theta_\ell+K_{16}\cos\theta_\ell\right)\cos\theta_b \cos\theta +  \nn\\
& \left(K_{17}\sin\theta_\ell\cos\theta_\ell+K_{18}\sin\theta_\ell\right)\sin\theta_b\cos\left(\phi_b+\phi_\ell\right)\cos\theta  +  \nn\\
& \left(K_{19}\sin\theta_\ell\cos\theta_\ell+K_{20}\sin\theta_\ell\right)\sin\theta_b\sin\left(\phi_b+\phi_\ell\right) \cos\theta +  \nn\\
& \left(K_{21}\cos\theta_\ell\sin\theta_\ell+K_{22}\sin\theta_\ell\right)\sin\phi_\ell \sin\theta +  \nn\\
& \left(K_{23}\cos\theta_\ell\sin\theta_\ell+K_{24}\sin\theta_\ell\right)\cos\phi_\ell  \sin\theta +  \nn\\
& \left(K_{25}\cos\theta_\ell\sin\theta_\ell+K_{26}\sin\theta_\ell\right)\sin\phi_\ell\cos\theta_b  \sin\theta +  \nn\\
& \left(K_{27}\cos\theta_\ell\sin\theta_\ell+K_{28}\sin\theta_\ell\right)\cos\phi_\ell\cos\theta_b  \sin\theta  +  \nn\\
& \left(K_{29}\cos^2\theta_\ell+K_{30}\sin^2\theta_\ell+{K_{35}\cos\theta_\ell}\right)\sin\theta_b\sin\phi_b  \sin\theta +  \nn\\
& \left(K_{31}\cos^2\theta_\ell+K_{32}\sin^2\theta_\ell+{K_{36}\cos\theta_\ell}\right)\sin\theta_b\cos\phi_b  \sin\theta +  \nn\\
& \left(K_{33}\sin^2\theta_\ell \right) \sin\theta_b\cos\left(2\phi_\ell+\phi_b\right) \sin\theta  +  \nn\\
& \left(K_{34}\sin^2\theta_\ell \right) \sin\theta_b\sin\left(2\phi_\ell+\phi_b\right)  \sin\theta  \Big).
\end{align} 

We identify the angular observables with those given \cite{Boer:2014kda} as: $K_{1ss} = K_1$, $K_{1cc} = K_2$, $K_{1c}=K_3$, $K_{2ss}=K_4$, $K_{2cc}=K_5$, $K_{2c}=K_6$, $K_{4sc}=K_7$, $K_{4s}=K_8$, $K_{3sc}=K_9$, and $K_{3s}=K_{10}$. We obtain two new angular coefficients $K_{35}$ and $K_{36}$ that are absent in the SM+SM$^\prime$ set of operators \cite{Blake:2017une}. These observables depend on the interference of scalar and (axial-)vector amplitudes only and are helicity suppressed by $m_\ell/\sqrt{q^2}$. Since we have retained the masses of the final state leptons, we write each of the $K_i$'s as
\begin{equation}\label{eq:Ki}
	K_{\{\cdots\}} = \mathcal{K}_{\{\cdots\}} + \frac{m_\ell}{\sqrt{q^2}} \mathcal{K}_{\{\cdots\}}^\prime + \frac{m_\ell^2}{q^2}\mathcal{K}_{\{\cdots\}}^{\prime\prime}\, .
\end{equation}
In Appendix \ref{sec:Ks} we express the observables in terms of transversity amplitudes. The massless part $\mathcal{K}_i$ of the observables $K_i$ have been previously calculated in the SM operator basis in \cite{Blake:2017une} and we agree with the results. Our expressions of $\mathcal{K}_i$ given in Appendix \ref{sec:Ks} extend the SM results by scalar amplitudes and are therefore new. The expressions of $\mathcal{K}^\prime_i$ and $\mathcal{K}^{\prime\prime}_i$ appear due to retaining the leptons masses and have not been calculated previously in the literature. If the final states leptons are of two lightest flavors, then $\mathcal{K}^{\prime,\prime\prime}$ can be neglect for large recoil analysis. But the expressions are useful for di-tau final states. The $\mathcal{K}^{\prime,\prime\prime}$ are also useful for accurate prediction of observables that are sensitive to lepton flavor universality violation.

Integrations \eqref{eq:angular} over the angles give differential decay distribution 
\begin{equation}
\frac{d\mathcal{B}}{dq^2} = 2K_1 + K_2\, .
\end{equation}
This is used to define normalized observables as
\begin{equation}
M_i = \frac{K_i}{d\mathcal{B}/dq^2}\, ,
\end{equation} 
The $M_i$'s can be extracted from \eqref{eq:angular} by convolution with different weight functions \cite{Blake:2017une}
\begin{align}
	M_i =  \frac{3}{32\pi^{2}} \bigintss\limits_{}^{}   \left( \sum\limits_{j=0}^{36} M_j f_{j}(\vec{\Omega}) \right) g_i (\vec\Omega) \deriv\vec{\Omega}
\end{align} 
where the weighting functions $g_i(\vec{\Omega})$ are chosen such that they satisfy
\begin{align}
	\int f_{j}(\vec{\Omega}) g_{i}(\vec{\Omega}) \deriv\vec{\Omega} = \left(\frac{32\pi^{2}}{3} \right) \delta_{ij}~.
\end{align} 
The weighting functions for $M_{1}$ to $M_{34}$ are given in \cite{Blake:2017une}. For $M_{35}$ and $M_{36}$ the weighting functions are
\begin{align}
g_{35}(\vec{\Omega}) &= \frac{9}{2}\cos\theta_\ell \sin\theta_b\sin\phi_b \sin\theta\, , \\
g_{36}(\vec{\Omega} )&= \frac{9}{2} \cos\theta_\ell \sin\theta_b\cos\phi_b  \sin\theta\, .
\end{align}

\section{Low-recoil Amplitudes in HQET \label{sec:HQET}}
At low recoil, lepton masses can be neglected for di-muon or di-electron final states. In that case, the time-like amplitudes $A_{\perp,\|t}$ and hence the form factor $f_t^{V,A}$ do not contribute. The HQET spin symmetry at leading order in $1/m_b$ expansion and up to $\ord(\alpha_s)$ corrections implies following relations between the rest of the form factors
\begin{align}\label{eq:LOIW}
& \xi_1-\xi_2 = f^V_\perp = f^V_0 = f^T_\perp = f^T_0\, ,\nn\\
& \xi_1+\xi_2 = f^A_\perp = f^A_0 = f^{T5}_\perp = f^{T5}_0\, ,
\end{align}
where $\xi_{1,2}$ are the leading Isgur-Wise form factors \cite{Feldmann:2011xf}. To exploit these relations we assume the vector form factors $f^{V,A}_{\perp,0}$ as independent and use the relations \eqref{eq:LOIW} for tensor form factors. Including one-loop corrections to the Isgur-Wise relations, the transversity amplitudes read \cite{Boer:2014kda}
\begin{align}\label{eq:HQETAmp1}
& A^{L(R)}_{\perp_1} \simeq -2N\mC_+^{L(R)} \sqrt{s_-} f_\perp^V\, ,\quad\quad A^{L(R)}_{\|_1} = 2N\mC^{L(R)}_-\sqrt{s_+} f^A_\perp \, ,\\
& A^{L(R)}_{\perp_0} \simeq \sqrt{2}N\mC_+^{L(R)} \frac{\mLb+\mL}{\sqrt{q^2}} \sqrt{s_-} f^V_0\, ,\\
\label{eq:HQETAmp2}
& A^{L(R)}_{\|_0} \simeq -\sqrt{2}N\mC_+^{L(R)} \frac{\mLb-\mL}{\sqrt{q^2}} \sqrt{s_+} f^A_0\, ,
\end{align}
where the Wilson coefficients are given by
\begin{align}
\begin{split}\label{eq:Cpm}
& \mC^{L(R)}_+ = \bigg( (\mC_9 + \mC_{9^\prime}) \mp (\mC_{10} + \mC_{10^\prime}) + \frac{2\kappa m_b \mLb}{q^2} \mC_7 \bigg)\, ,\\
&\mC^{L(R)}_- = \bigg( (\mC_9 - \mC_{9^\prime}) \mp (\mC_{10} - \mC_{10^\prime}) + \frac{2\kappa m_b \mLb}{q^2} \mC_7  \bigg)\, .
\end{split}
\end{align}
The parameter $\kappa \equiv \kappa(\mu) = 1 - (\alpha_sC_F/2\pi)\ln(\mu/m_b)$ accounts for the radiative QCD corrections to the form factors relations. The parameter $\kappa$ is such that together with the Wilson coefficients and the $\overline{\rm MS}$ $b$-quark mass $m_b$ in \eqref{eq:Cpm}, the amplitudes are free of the renormalization scale $\mu$.

The simplifications of the transversity amplitudes yield factorizations between short- and long-distance physics in the angular observables. In the factorized expressions, the vector and axial-vector Wilson coefficient contributes through the following short-distance coefficients \cite{Boer:2014kda}
%
%
\begin{align}
\begin{split}
\rho_1^\pm &= \frac{1}{2} \left( |\mC^{R}_{\pm}|^2 + |\mC^L_{\pm}|^2 \right) = |\mC_{79} \pm \mC_{9'}|^2 + |\mC_{10} \pm \mC_{10'}|^2 \,, \cr
\rho_2 &= \frac{1}{4} \left( \mC^{R}_{+} \mC^{R*}_{-} - \mC^{L}_{-} \mC^{L*}_{+} \right) = \re ({\mC_{79} \mC^{*}_{10} - \mC_{9'} \mC^{*}_{10'}}) - i \im ({\mC_{79} \mC^{*}_{9'} + \mC_{10} \mC^{*}_{10'}}) \,.\cr
%
\rho_3^\pm &= \frac{1}{2} \left( |\mC^{R}_{\pm}|^2 - |\mC^{L}_{\pm}|^2 \right)= 2 \, \re{(\mC_{79} \pm \mC_{9'}) (\mC_{10} \pm \mC_{10'})^*} \\
\rho_4 &= \frac{1}{4} \left( \mC^{R}_{+} \mC^{R*}_{-} + \mC^{L}_{-} \mC^{L*}_{+} \right) \cr &= \left( |\mC_{79}|^2 - |\mC_{9'}|^2 + |\mC_{10}|^2 - |\mC_{10'}|^2 \right) - i\im{\mC_{79} \, \mC^{*}_{10'} - \mC_{9'}\, \mC^{*}_{10}} \,,
\end{split}
\end{align}
where we have abbreviated
\begin{equation}
\begin{aligned}
\mC_{79}   & \equiv \mC_{9} + \frac{2\kappa m_b \mLb}{q^2} \mC_{7}\,. &\end{aligned}
\end{equation}
The $\rho_3^\pm$ and $\rho_4$ contributes due to the secondary parity violating decay. These coefficients appear in other $b\to s\ell^+\ell^-$ decays also. For example, $\rho^\pm_1$ and $\rho_2$ appear in $B\to K^\ast(\to K\pi)\ell^+\ell^-$ \cite{Bobeth:2010wg}, $B\to K\pi\ell^+\ell^-$ \cite{Das:2014sra}, and $\Lambda_b\to\Lambda^\ast(\to N\!\bar{K})\ell^+\ell^-$ decay \cite{Das:2020cpv}, $\rho_4$ appears in $\Lambda_b\to\Lambda^\ast(\to N\!\bar{K})\ell^+\ell^-$ decay, and $\rho_3^-$ and $\rho_4$ appear in $B\to K\pi\ell^+\ell^-$ decay. 
In the SM the following simplifications take place 
\begin{align}
&\rho_1^+ = \rho_1^- = \rho_1 = 2 \re(\rho_4)\,, \quad \rho_3^+ = \rho_3^- = \rho_3\,,\\
&\im(\rho_2) = 0\, ,\quad \im(\rho_4) = 0\, ,
\end{align}
so that the independent coefficients are $\rho_1, \rho_3$ and $\re(\rho_2)$. The scalar short-distance coefficients that appear in angular observables are
\begin{align}\begin{split}
&\rho_{\rm S}^\pm=|\mC_{\rm SP\pm}^{L}|^2 + |\mC_{\rm SP\pm}^{R}|^2\, , \\
&\rho_{\rm S1}= 2(\mC_{\rm SP+}^{L} \mC_{\rm SP-}^{L*} + \mC_{\rm SP+}^{R} \mC_{\rm SP-}^{R*}  )\, .
\end{split}
\end{align}

\section{Low-recoil Factorization \label{sec:lowrecoil}}
Using the HQET simplifications of the previous section, we obtain factorization between short- and long-distance physics in the angular observables. We reiterate that the lepton mass is neglected to obtain these expressions.  The factorized expression for the 10 observables that also appear in the unpolarized $\Lambda_b$ decay are
\begin{align} 
K_{1} 
& = 2 s_{+} \left( |f_{\perp}^{A}|^{2} + \frac{(m_{\Lb} - m_{\Lz})^{2}}{q^{2}} |f_{0}^{A}|^{2} \right) \rho_{1}^{-} 
+ 2 s_{-} \left( |f_{\perp}^{V}|^{2} + \frac{(m_{\Lb} + m_{\Lz})^{2}}{q^{2}} |f_{0}^{V}|^{2}  \right) \rho_{1}^{+} \nn\\
&\quad+\frac{1}{m_b^2}[|f_{t}^{V}|^{2}  s_{+} \rho_{S+}(\mLb - \mL)^2 +|f_{t}^{A}|^{2} s_{-} \rho_{S-}(\mLb + \mL)^2 ]~,\\ 
K_{2} & = 4 s_{+} |f_{\perp}^{A}|^{2} \rho_{1}^{-} + 4 s_{-} |f_{\perp}^{V}|^{2} \rho_{1}^{+} + 
\frac{1}{m_b^2}[|f_{t}^{V}|^{2}  s_{+} \rho_{S+}(\mLb - \mL)^2 \nn\\&\quad+|f_{t}^{A}|^{2} s_{-} \rho_{S-}(\mLb + \mL)^2 ]~, \\
K_{3} & = 16 \sqrt{s_+ s_-} f_{\perp}^{A} f_{\perp}^{V} {\rm Re}(\rho_2) ~, \\
K_{4} & = -8 \alpha_{\Lz} \sqrt{s_{+} s_{-}} 
\left( f_{\perp}^{A} f_{\perp}^{V}  + \frac{(m_{\Lb}^{2} - m_{\Lz}^{2})}{q^{2}} f_{0}^{V} f_{0}^{A} \right) {\rm Re}(\rho_{4}) \nn\\
&- \alpha_{\Lambda}  \sqrt{s_{+}s_{-}} f_t^V f_t^A (\mLb^2 - \mL^2)\re(\rho_{S1}) ~, \\
K_{5} & = -16 \alpha_{\Lz} \sqrt{s_{+} s_{-}} f_{\perp}^{A} f_{\perp}^{V} {\rm Re}(\rho_{4})
- \alpha_{\Lambda} \sqrt{s_{+}s_{-}} f_t^V f_t^A  (\mLb^2 - \mL^2) \re(\rho_{S1})~, \\
K_{6} & = 
-4 \alpha_{\Lz} s_{+} |f_{\perp}^{A}|^{2} \rho_{3}^{-} 
- 4 \alpha_{\Lz} s_{-} |f_{\perp}^{V}|^{2} \rho_{3}^{+} 
~, \\
K_{7} &=  -8 \alpha_{\Lz} \sqrt{s_{+}s_{-}} \left( 
\frac{(m_{\Lb} + m_{\Lz})}{\sqrt{q^2}} f_{0}^{V} f_{\perp}^{A} - 
\frac{(m_{\Lb} - m_{\Lz})}{\sqrt{q^2}} f_{0}^{A} f_{\perp}^{V}
\right) {\rm Re}(\rho_{4})~\\
K_{8} & = 
4 s_{+} \alpha_{\Lz} \frac{(m_{\Lb} - m_{\Lz})}{\sqrt{q^2}} f_{0}^{A} f_{\perp}^{A} \rho_{3}^{-} -  
4 s_{-} \alpha_{\Lz} \frac{(m_{\Lb} + m_{\Lz})}{\sqrt{q^2}} f_{0}^{V} f_{\perp}^{V} \rho_{3}^{+} 
~, \\
K_{10} & = 8 \alpha_{\Lz} \sqrt{s_+ s_-} \left(
\frac{(m_{\Lb} + m_{\Lz})}{\sqrt{q^2}} f_{0}^{V} f_{\perp}^{A}  + 
\frac{(m_{\Lb} - m_{\Lz})}{\sqrt{q^2}}  f_{0}^{A} f_{\perp}^{V}
\right) {\rm Im}(\rho_{4})~, 
\end{align} 
For the additional observables that appear due to polarization, the factorization looks like 
\begin{align}
%
K_{11} & = -8 P_{\Lb} \sqrt{s_+ s_-}  \left( 
f_0^A f_0^V \frac{(m_{\Lb}^2 - m_{\Lz}^2)}{q^2} - f_\perp^A f_\perp^V 
\right) {\rm Re}(\rho_4)\nn\\
&-P_{\Lambda_b} N^2 \alpha_{\Lambda} f_t^V f_t^A \re[\rho_{S1}] \sqrt{s_{+}s_{-}}  (\mLb^2 - \mL^2)  ~,   \\
K_{12} &= 16 P_{\Lb} \sqrt{s_+ s_-} f_\perp^A f_\perp^V {\rm Re}(\rho_4) 
-P_{\Lambda_b} N^2 \alpha_{\Lambda} f_t^V f_t^A \re[\rho_{S1}] \sqrt{s_{+}s_{-}}  (\mLb^2 - \mL^2) ~, \\
K_{13} &= 
 4 P_{\Lb} s_ {+} |f_{\perp}^{V}|^{2} \rho_{3}^{-} +
4 P_{\Lb} s_{-} |f_{\perp}^{A}|^{2} \rho_{3}^{+} ~,  \\
K_{14} &= -2 \alpha_{\Lz} P_{\Lb} s_- \left( |f_{\perp}^{V}|^{2} - |f_{0}^{V}|^{2} \frac{(m_{\Lb} +  m_{\Lz})^{2}}{q^2}\right) \rho_{1}^{+} \nn\\&\quad - 2 \alpha_{\Lz} P_{\Lb}  s_+ \left( |f_{\perp}^{A}|^{2} - |f_{0}^{A}|^{2}  \frac{(m_{\Lb} -  m_{\Lz})^{2}}{q^2} \right) \rho_{1}^{-} \\
&+P_{\Lambda_b} \alpha_{\Lambda} N^2 \frac{1}{m_b^2}[|f_{t}^{V}|^{2}  s_{+} \rho_{S+}(\mLb - \mL)^2 +|f_{t}^{A}|^{2} s_{-} \rho_{S-}(\mLb + \mL)^2 ] ~,\\ 
K_{15} &= 
-4 \alpha_{\Lz} P_{\Lb} s_{-} |f_{\perp}^V|^2 \rho_1^{+}    
-4 \alpha_{\Lz} P_{\Lb} s_{+}|f_{\perp}^A|^2  \rho_1^{-}\nn\\
&+P_{\Lambda_b} \alpha_{\Lambda} N^2 \frac{1}{m_b^2}[|f_{t}^{V}|^{2}  s_{+} \rho_{S+}(\mLb - \mL)^2 +|f_{t}^{A}|^{2} s_{-} \rho_{S-}(\mLb + \mL)^2 ] ~ , \\
K_{16} &= -16 \alpha_{\Lz} P_{\Lb}\sqrt{s_+ s_-} f_{\perp}^A f_{\perp}^V {\rm Re}(\rho_2)~,  \\
K_{17} &= 
-4 \alpha_{\Lz} P_{\Lb}  s_{-} \frac{(m_{\Lb} + m_{\Lz})}{\sqrt{q^2}} f_{0}^{V} f_{\perp}^{V} \rho_{1}^{+} 
+4 \alpha_{\Lz} P_{\Lb} s_{+} \frac{(m_{\Lb} - m_{\Lz})}{\sqrt{q^2}} f_{0}^{A} f_{\perp}^{A} \rho_{1}^{-} ~,\\
K_{18} &= -16 \alpha_{\Lz} P_{\Lb} \sqrt{s_+ s_-} \left(
\frac{(m_{\Lb} + m_{\Lz})}{\sqrt{q^2}} f_0^V f_\perp^A  -  
\frac{(m_{\Lb} - m_{\Lz} )}{\sqrt{q^2}} f_0^A f_\perp^V 
\right) {\rm Re}(\rho_2)~,
\end{align} 
\begin{align} 
K_{19} &= 8 \alpha_{\Lz} P_{\Lb} \sqrt{s_+ s_-} \left(
\frac{(m_{\Lb} + m_{\Lz})}{\sqrt{q^2}} f_0^V f_\perp^A  + 
\frac{(m_{\Lb} - m_{\Lz} )}{\sqrt{q^2}} f_0^A f_\perp^V 
\right) {\rm Im}(\rho_2) ~,\\
K_{22} &= -8 P_{\Lb} \sqrt{s_+ s_-} \left(
\frac{(m_{\Lb} + m_{\Lz})}{\sqrt{q^2}} f_{0}^{V} f_{\perp}^{A}  - 
\frac{(m_{\Lb} - m_{\Lz} )}{\sqrt{q^2}} f_{0}^{A} f_{\perp}^{V} 
\right) {\rm Im}(\rho_4) ~,\\
K_{23} &= 8 P_{\Lb} \sqrt{s_+ s_-}\left( 
\frac{(m_{\Lb} + m_{\Lz})}{\sqrt{q^2}} f_0^V f_\perp^A  + 
\frac{(m_{\Lb} - m_{\Lz} )}{\sqrt{q^2}} f_0^A f_\perp^V 
\right) {\rm Re}(\rho_4) ~,\\
K_{24} &= 4 P_{\Lb}  s_{-} \frac{(m_{\Lb} + m_{\Lz})}{\sqrt{q^2}} f_{0}^{V} f_{\perp}^{V} \rho_{3}^{+} 
+4 P_{\Lb} s_{+} \frac{(m_{\Lb} - m_{\Lz})}{\sqrt{q^2}} f_{0}^{A} f_{\perp}^{A} \rho_{3}^{-} ~,\\
K_{25} &= 8 \alpha_{\Lz} P_{\Lb} \sqrt{s_+ s_-} \left( 
\frac{(m_{\Lb} + m_{\Lz})}{\sqrt{q^2}} f_0^V f_{\perp}^{A} -  
\frac{(m_{\Lb} - m_{\Lz})}{\sqrt{q^2}} f_0^A f_{\perp}^{V} 
\right)  {\rm Im}(\rho_2) ~,\\
K_{27} &= 
-4 \alpha_{\Lz} P_{\Lb} s_{-} \frac{(m_{\Lb} + m_{\Lz})}{\sqrt{q^2}} f_0^V f_{\perp}^{V} \rho_1^+ - 
4 \alpha_{\Lz} P_{\Lb} s_{+} \frac{(m_{\Lb} - m_{\Lz})}{\sqrt{q^2}} f_0^A f_{\perp}^{A} \rho_1^- ~, \\ 
K_{28} &=- 8 \alpha_{\Lz} P_{\Lb} \sqrt{s_+ s_-} \left( 
\frac{(m_{\Lb} + m_{\Lz})}{\sqrt{q^2}} f_{0}^{V} f_{\perp}^{A} + 
\frac{(m_{\Lb} - m_{\Lz})}{\sqrt{q^2}} f_{0}^{A} f_{\perp}^{V} 
\right) {\rm Re}(\rho_2)  ~, \\
\label{eq:K29}
K_{29}&=-P_{\Lambda_b}  \alpha_{\Lambda} f_t^V f_t^A \im(\rho_{S1}) \sqrt{s_{+}s_{-}}  (\mLb^2 - \mL^2) \\
K_{30} &= +8 \alpha_{\Lz} P_{\Lb} \sqrt{s_+ s_-}\frac{( m_{\Lb}^2 - m_{\Lz}^2 )}{q^2} f_0^A f_0^V {\rm Im}(\rho_2)
\nn\\&\quad+P_{\Lambda_b} \alpha_{\Lambda} f_t^V f_t^A \im(\rho_{S1}) \sqrt{s_{+}s_{-}}  (\mLb^2 - \mL^2) 
\\
\label{eq:K31}
K_{31} &=P_{\Lambda_b} \alpha_{\Lambda} \frac{1}{m_b^2}[-|f_{t}^{V}|^{2}  s_{+} ~\rho_{S+}~(\mLb - \mL)^2 +|f_{t}^{A}|^{2} s_{-} ~\rho_{S-}~(\mLb + \mL)^2 ] \\
K_{32} &= -2 \alpha_{\Lz} P_{\Lb} s_-\frac{( m_{\Lb} + m_{\Lz} )^2}{q^2}|f_0^V|^2\rho_1^+ +
2 \alpha_{\Lz} P_{\Lb} s_+\frac{( m_{\Lb} - m_{\Lz} )^2}{q^2}|f_0^A|^2\rho_1^- \\
&+ \alpha_{\Lambda}P_{\Lambda_b} \frac{1}{m_b^2}[-|f_{t}^{V}|^{2}  s_{+} \rho_{S+}(\mLb - \mL)^2 +|f_{t}^{A}|^{2} s_{-} \rho_{S-}(\mLb + \mL)^2 ] ~, \\ 
K_{33} &= -2 \alpha_{\Lz} P_{\Lb}  s_- |f_{\perp}^{V}|^{2}  \rho_1^+ 
+  2 \alpha_{\Lz} P_{\Lb}s_+ |f_{\perp}^{A}|^{2}  \rho_1^- ~,\\
K_{34} &= 8 \alpha_{\Lz} P_{\Lb} \sqrt{ s_+ s_- } f_{\perp}^{A} f_{\perp}^{V} {\rm Im}(\rho_2) ~. 
%
\label{eq:rhodependence}
\end{align}
Note that the observables $K_{20,21}$ vanish in the HQET. Moreover, $K_{35}$ and $K_{36}$ are helicity suppressed by $m_\ell/\sqrt{q^2}$ and therefore are not given above.
 
Using the factorized expressions, we derive tests of operator product expansion through form factors and the short-distance physics. In the SM+SM$^\prime$+SP set of operators, the combinations of $K_2$ and $K_{33}$ are 
\begin{align}
& K_2 - \frac{2}{\alpha_\Lambda P_{\Lambda_b}} K_{33} = |N|^2(8 |f^A_\perp|^2 s_+ \rho_1^- + |f^V_t|^2 s_+ \frac{(\mLb-\mL)^2}{m_b^2} \rho_{\rm S}^+ + |f^A_t|^2 s_- \frac{(\mLb+\mL)^2}{m_b^2} \rho_{\rm S}^-)\, ,\\
& K_2 + \frac{2}{\alpha_\Lambda P_{\Lambda_b}} K_{33} = |N|^2 (8 |f^V_\perp|^2 s_- \rho_1^+ + |f^V_t|^2s_+ \frac{(\mLb-\mL)^2}{m_b^2} \rho_{\rm S}^+ + |f^A_t|^2 s_- \frac{(\mLb+\mL)^2}{m_b^2} \rho_{\rm S}^-)\, .
\end{align}
If the SP operators are absent then these combinations depend only on $\rho^-$ and $\rho^+$ respectively. The $K_8$ and $K_{24}$ can be combined to form two relations that can be used to extract $\rho_3^+$ and $\rho_3^-$, and $K_{17}$ and $K_{27}$ can be combined to form two relations that can be used to extract $\rho^\pm_1$ even in the presence of SP operators
\begin{align}
& P_{\Lambda_b} K_8 + \alpha_\Lambda K_{24} = -8|N|^2 P_\Lb \alpha_\Lam f^V_0 f^V_\perp \frac{\mLb+\mL}{\sqrt{q^2}} s_- \rho_3^+ \, ,\\
& P_{\Lambda_b} K_8 - \alpha_\Lambda K_{24} = 8|N|^2 P_\Lb \alpha_\Lam f^A_0 f^A_\perp \frac{\mLb-\mL}{\sqrt{q^2}} s_+ \rho_3^-\, , \\
& K_{27} + K_{17} = 8|N|^2 P_\Lb \alpha_\Lam f^A_0 f^A_\perp \frac{\mLb-\mL}{\sqrt{q^2}} s_+ \rho_1^-\, ,\\
&K_{27} - K_{17} = -8 |N|^2 P_\Lb \alpha_\Lam f^V_0 f^V_\perp \frac{\mLb+\mL}{\sqrt{q^2}} s_- \rho_1^+\, .
\end{align}

In the SM+SM$^\prime$+SP set of operators, several ratios of short-distance coefficients can be determined without any hadronic effects up to $\Lambda_{\rm QCD}/m_b$ corrections
\begin{align}
&\frac{P_{\Lambda_b} K_8 + \alpha_\Lambda K_{24}}{K_{27} - K_{17}} = - \frac{\rho_3^-}{\rho_1^-}\, ,\quad \frac{P_{\Lambda_b} K_8 - \alpha_\Lambda K_{24}}{K_{27} + K_{17}} =  \frac{\rho_3^+}{\rho_1^+}\, ,
\end{align}
\begin{align}
\frac{K_{16}}{K_{34}} =- \frac{2\re(\rho_2)}{\im(\rho_2)}\, ,\quad \frac{K_{25}}{K_{22}} = -\frac{\alpha_\Lambda\im(\rho_2)}{\im(\rho_4)}\, ,\quad \frac{K_{23}}{K_{10}} = -\frac{P_{\Lambda_b} \re(\rho_4)}{\alpha_\Lambda \im(\rho_4)}\, .
\end{align}

In the SM+SM$^\prime$ the ratio of $K_3$ and $K_5$ is proportional to $\re(\rho_2)/\re(\rho_4)$ but is modified as following in the presence of SP operators
\begin{equation}\label{eq:K53relation}
\frac{K_3}{K_5} = -\frac{16 m_b^2 f^A_\perp f^V_\perp \re(\rho_2) }{16 \alpha_\Lambda m_b^2 f^A_\perp f^V_\perp \re(\rho_4) + \alpha_\Lambda(\mmLb-\mmL) f^A_t f^V_t \re(\rho_{\rm S1}) }\, .
\end{equation}
If the SP operators are absent, then this ratio is equal to  $\re(\rho_2)/\alpha_\Lambda\re(\rho_4)$. On the other hand, in SM+SM$^\prime$, the ratios $K_5/K_7$ and $K_{5}/K_{23}$ are independent of any short distance physics but are modified in the presence of SP operators as
\begin{align}
\label{eq:K57relation}
&\frac{K_5}{K_7} = \frac{\sqrt{q^2}\bigg[32 f^A_\perp f^V_\perp \re(\rho_4) + (\mmLb-\mmL)f^A_t f^V_t \re(\rho_{\rm S1}) \bigg]}{16 m_b^2\re(\rho_4)\bigg[ (\mLb+\mL)f^A_\perp f^V_0  - (\mLb-\mL)f^A_0 f^V_\perp \bigg]}\, ,\\
\label{eq:K523relation}
&\frac{K_{5}}{K_{23}} = \frac{\sqrt{q^2} \bigg[32 f^A_\perp f^V_\perp \re(\rho_4) + (\mmLb-\mmL)f^A_t f^V_t \re(\rho_{\rm S1}) \bigg] }{16 m_b^2 P_{\Lambda_b} \re(\rho_4) \bigg[ (\mLb+\mLb) f^A_\perp f^V_0 + (\mLb-\mL) f^A_0 f^V_\perp \bigg] } \, .
\end{align}
These two relations could be regarded as null test of SM+SM$^\prime$.

We find a ratios involving $K_{18}$, $K_{28}$, and $K_3$ that are independent of any short distance physics in the SM+SM$^\prime$+SP set of operators  
\begin{align}
&\frac{K_{18} + K_{28}}{K_3} = -P_{\Lambda_b}\alpha_\Lambda \frac{\mLb+\mL}{\sqrt{q^2}} \frac{f^V_0}{f^V_\perp}\, ,\\
&\frac{K_{18} - K_{28}}{K_3} = P_{\Lambda_b}\alpha_\Lambda \frac{\mLb-\mL}{\sqrt{q^2}} \frac{f^A_0}{f^A_\perp}\, .
\end{align}
The ratios can be used to extract $f^A_0/f^A_\perp$ and $f^V_0/f^V_\perp$.

The angular observable $K_{29}$ vanishes in the SM+SM$^\prime$ in the limit of zero lepton mass. If SP operators are present then it is proportional to the imaginary part of $\rho_{\rm S1}$ as given in equation \eqref{eq:K29}. The real part of $\rho_{\rm S1}$ can be extracted from the combination
\begin{equation}\label{combo:rhoS1}
P_{\Lambda_b} (K_4-K_5) - \alpha_\Lambda K_{11} = P_{\Lambda_b} \alpha_\Lambda f^A_t f^V_t \frac{(\mmLb-\mmL)}{m_b^2} \sqrt{s_+s_-} \re(\rho_{\rm S1})\, , 
\end{equation}
so that the $\rho_{\rm S1}$ can be determined completely.

$K_{31}$ given in equation \eqref{eq:K31} also vanishes in the SM if $m_\ell$ is neglected but depends on $\rho_{\rm S\pm}$ if scalar operators are present. This observable therefore serves as a null test of SM. We find another null test of $\rho_{\rm S\pm}$ from the following combination
\begin{equation}\label{eq:K215relation}
K_2 + \frac{K_{15}}{P_{\Lambda_b}\alpha_\Lambda} = 2\bigg( |f^V_t|^2 s_+\frac{(\mLb-\mL)^2}{m_b^2} \rho_{\rm S}^+ + |f^A_t|^2 s_-\frac{(\mLb+\mL)^2}{m_b^2} \rho_{\rm S}^- \bigg)\, .
\end{equation}
This equation in combination with \eqref{eq:K31} can be used to extract $\rho_S^+$ and $\rho_S^-$. We find that the following relation holds in the presence of SM+SM$^\prime$+SP set of operators
\begin{equation}
P_{\Lambda_b} \alpha_\Lambda K_2 - K_{15} = 2 \bigg( P_{\Lambda_b} \alpha_\Lambda K_1 - K_{14} \bigg) \,.
\end{equation}

Since we work in the $m_\ell\to 0$ limit, the relations $K_{13}\alpha_\Lambda=-P_{\Lambda_b} K_6$ and $K_{16}=-P_{\Lambda_b}\alpha_\Lambda K_3$ remain unchanged when SM+SM$^\prime$ is supplemented by the SP operators \cite{Blake:2017une} in low-recoil HQET.

\section{Numerical Analysis \label{sec:numerical}}
In this section we perform a numerical analysis of polarized $\Lambda_b$ decay to $\Lambda(\to p\pi)\mu^+\mu^-$ at the low recoil region $15\gev^2\le q^2\le 20\gev^2$. At this region the masses of the leptons are neglected. Due to low-recoil HQET, only four form factors contribute through the SM amplitudes \eqref{eq:HQETAmp1}-\eqref{eq:HQETAmp2} and the scalar amplitudes \eqref{TA:scalar1} and \eqref{TA:scalar2}. The form factors, as well as their uncertainties are taken from lattice QCD calculations \cite{Detmold:2016pkz}. In addition to uncertainties coming from the form factors, there are additional sources of uncertainties in our analysis. We consider a 10\% corrections between the amplitudes \eqref{eq:HQETAmp1}-\eqref{eq:HQETAmp2} to account for the neglected terms of the $\mathcal{O}(\Lambda_{\rm QCD}/m_b)$ and $\mathcal{O}(\mL/\mLb)$ to derive the leading Isgur-Wise relations \eqref{eq:LOIW}. These corrections are implemented by scaling the amplitudes $A_{\perp,\|0}$ by uncorrelated real scale factors.

There are theoretical uncertainties emanating from a virtual photon connected to the operators $\mathcal{O}_{1-6}$ and $\mathcal{O}_8$ which are of nonlocal in nature. At low recoil, the HQET combined with low recoil OPE in $1/Q$, where $Q\sim (m_b,\sqrt{q^2})$, the nonlocal effects are calculated in terms of local matrix elements suppressed by the powers of $Q$ and absorbed in the Wilson coefficients $\mC_{7,9}$ \cite{Grinstein:2004vb}. The uncertainties due to the neglected terms in the OPE of the order $\mathcal{O}(\alpha_s\Lambda_{\rm QCD}/m_b, m_c^4/Q^4)$ are included in our analysis by a uncorrelated 5\% corrections to the amplitudes $A_{\perp,\|0}^{L(R)}$. Other sources of uncertainties are parametric uncertainties, and scale dependence of the SM Wilson coefficients $\mC_i(\mu)$ for varying the scale $\mu$ in the range $m_b/2<\mu<2m_b$. At large-$q^2$, uncertainties due to quark-hadron duality violation in the integrated observables are expected to be small and are neglected in our analysis \cite{Beylich:2011aq}. 

\begin{figure}[h!]
	\begin{center}
		\includegraphics[scale=0.41]{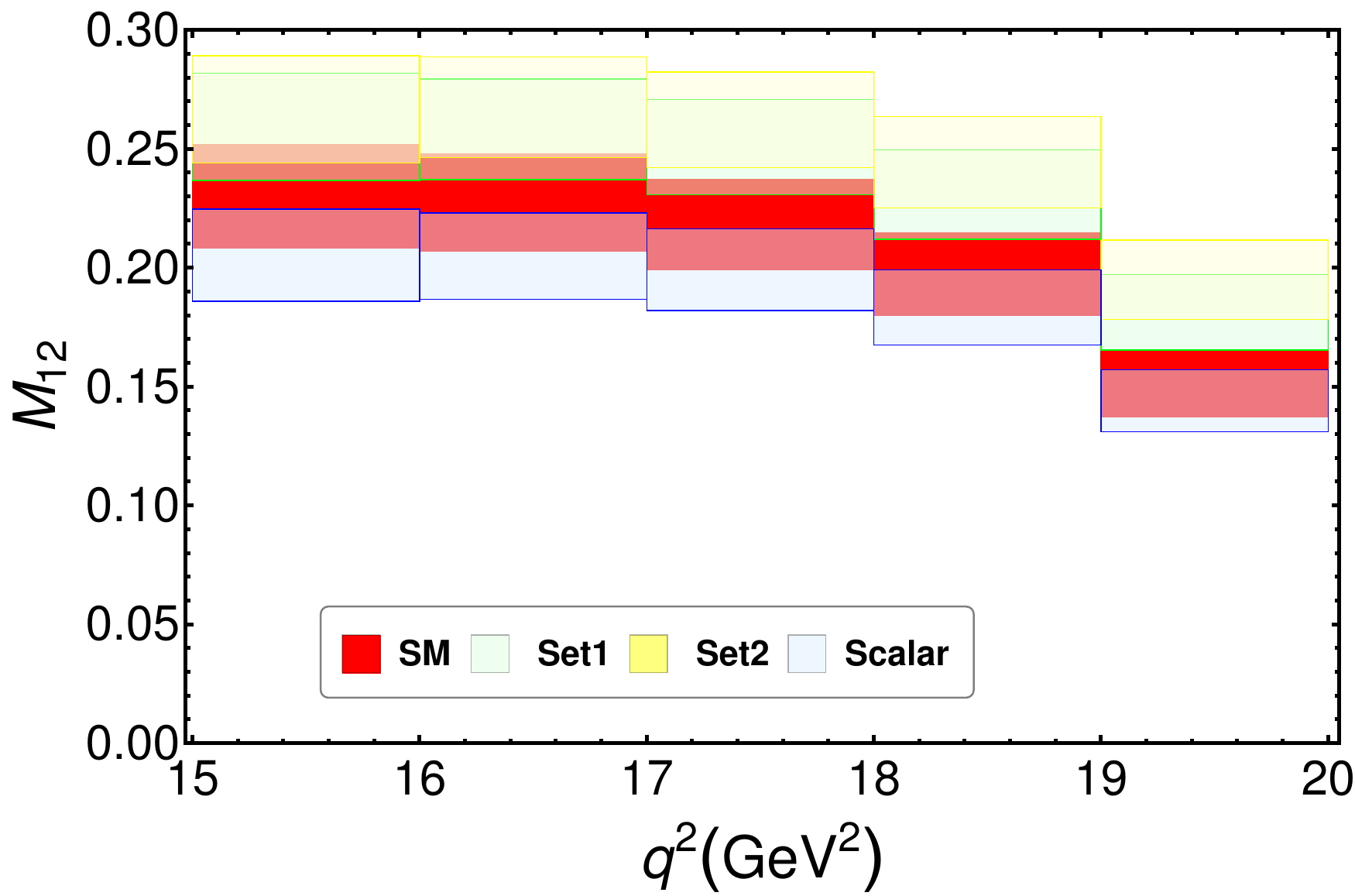}
		\includegraphics[scale=0.4]{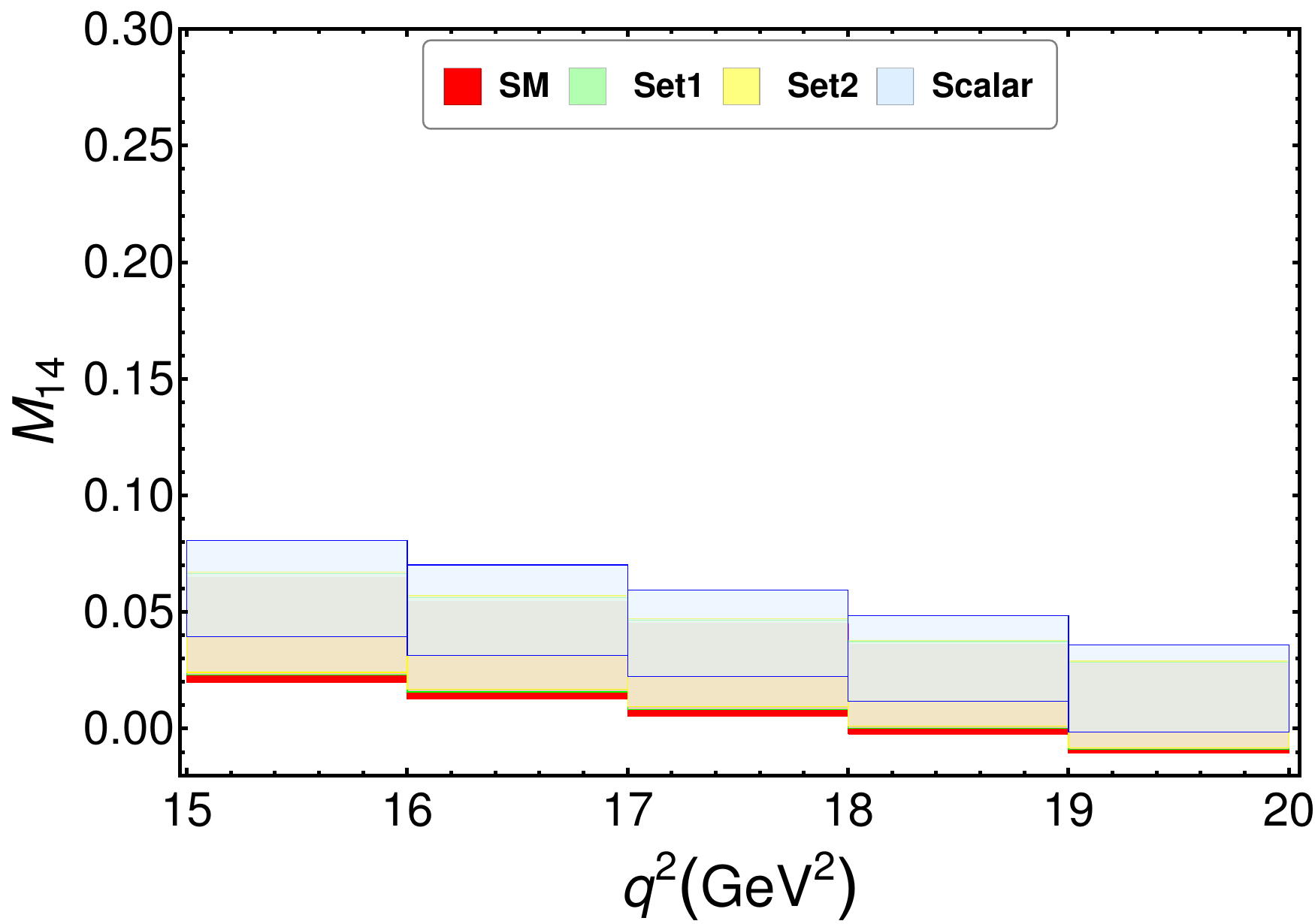}
		\includegraphics[scale=0.4]{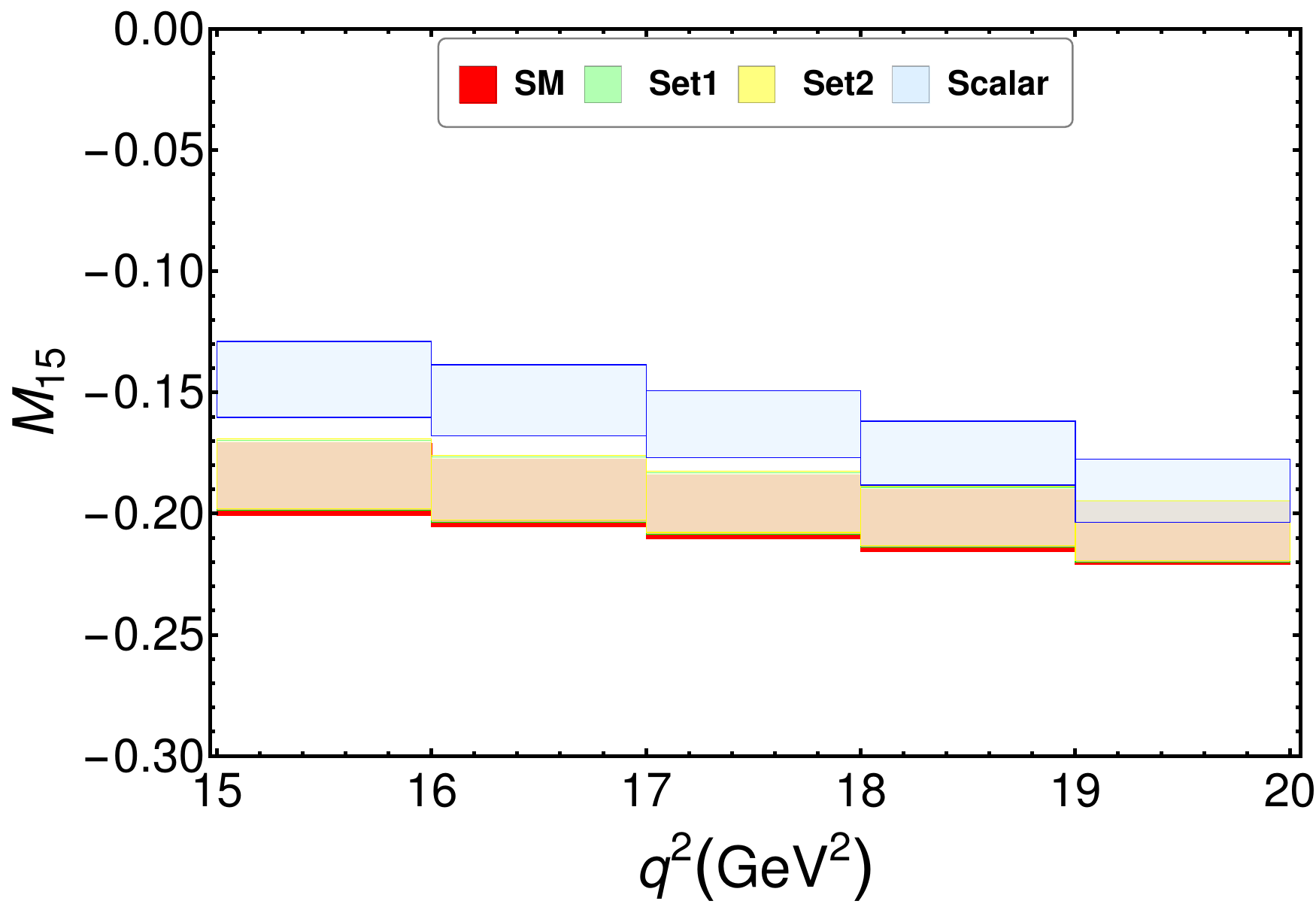}
		\includegraphics[scale=0.4]{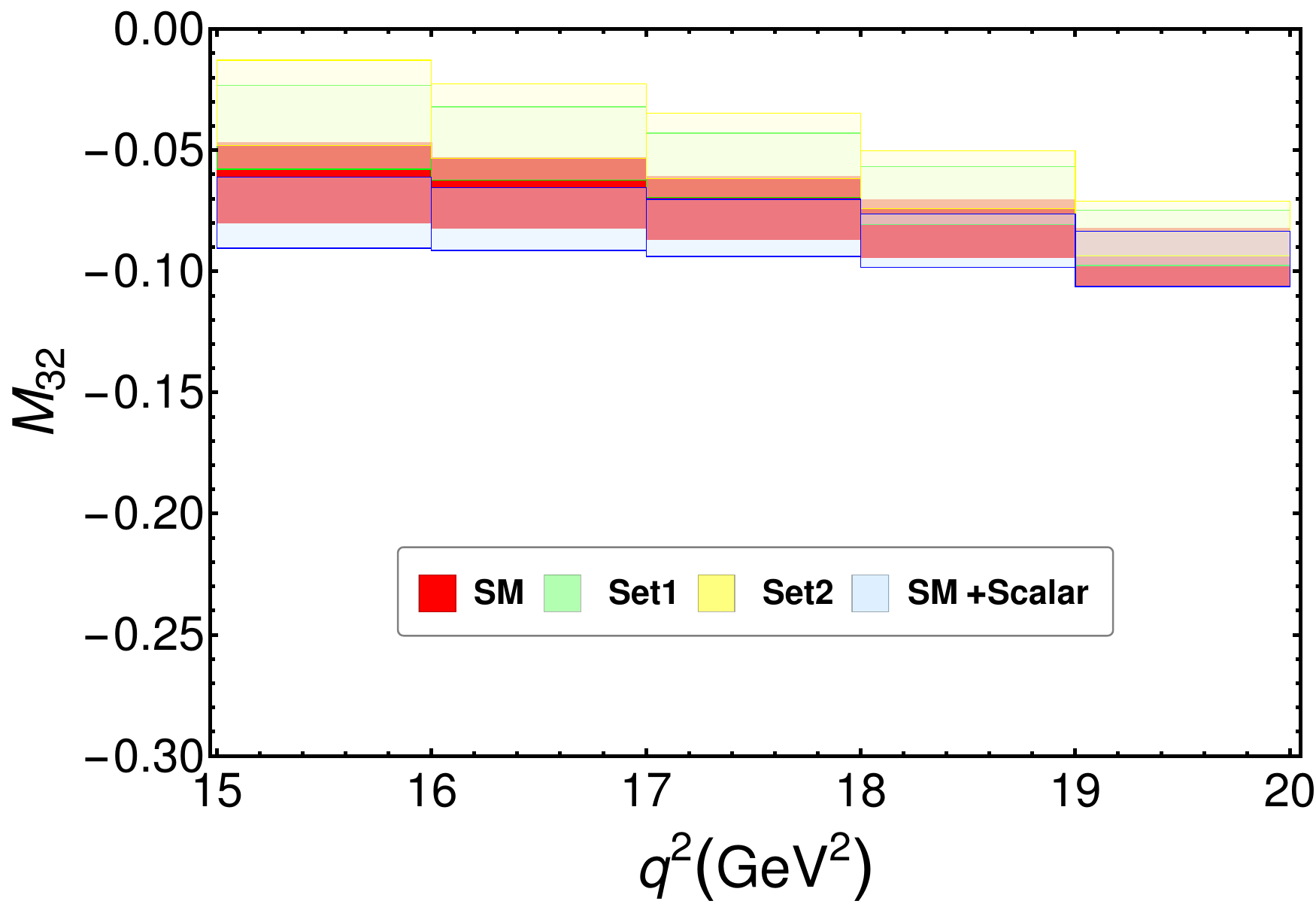}
		\caption{The angular observables $M_{12, 14, 15, 32}$ for polarized $\Lambda_b\to\Lambda(\to p\pi)\mu^+\mu^-$ decay in different $q^2$ bins in the SM and NP scenarios. For (axial-)vector operators we choose two sets from global fits to $b \to s\mu^+\mu^-$ data. Set1 corresponds to set $\delta\mC_9 = -\mC_9^\prime = -1.21\, ,\quad \delta\mC_{10} = +\mC_{10}^\prime = 0.28$ and Set2 corresponds to $\delta\mC_9 = -\mC_9^\prime = -1.11\, ,\quad \delta\mC_{10} = -\mC_{10}^\prime = 0.09$. The set of scalar couplings are $\mC_S = 1.5 + i  0.03, \mC_S^{\prime} = -2 + i 0.2, \mC_P = 1.8 + i 0.1, \mC_P^{\prime} = -1.1-i 0.04 $.
			\label{fig:NPplots1}}
	\end{center} 
\end{figure}

Though there are a total of 36 observables, for simplicity we study the ones that otherwise do not appear in unpolarized $\Lambda_b$ decay, and where both SP and (axial-)vector operators contribute. We also show in our plots only the interesting variants of observables for a representative set of NP couplings. A detailed NP analysis for all the observables will be presented elsewhere. To estimate the effects of NP operators we summarize the constraints on the NP Wilson coefficients that we use for our analysis. For simplicity we assume that the vector and axial vector, and scalar and pseudoscalar NP contributes separately. From the global fits to $b\to s\mu^+\mu^-$ data we we choose two sets for illustrations \cite{Capdevila:2018jhy,Capdevila:2017bsm,Blake:2019guk,Altmannshofer:2021qrr,Alguero:2021anc,Geng:2021nhg}, Set1: $\delta\mC_9 = -\mC_9^\prime = -1.21\, ,\quad \delta\mC_{10} = +\mC_{10}^\prime = 0.28$ 
and       Set2: $\delta\mC_9 = -\mC_9^\prime = -1.11\, ,\quad \delta\mC_{10} = -\mC_{10}^\prime = 0.09$. For scalar Wilson coefficients we follow the fit presented in \cite{Beaujean:2015gba}. 

All our determination are for $P_{\Lambda_b}=1$. In figure \ref{fig:NPplots1} we show the SM predictions and NP sensitivities of $M_{11}$, $M_{12}$, $M_{14}$, $M_{32}$ different high-$q^2$ bins. Here the bands correspond to the uncertainties coming from form factor and other sources. These plots indicate that the observables are sensitive to NP effects.

\begin{figure}[h!]
	\begin{center}
		\includegraphics[scale=0.4]{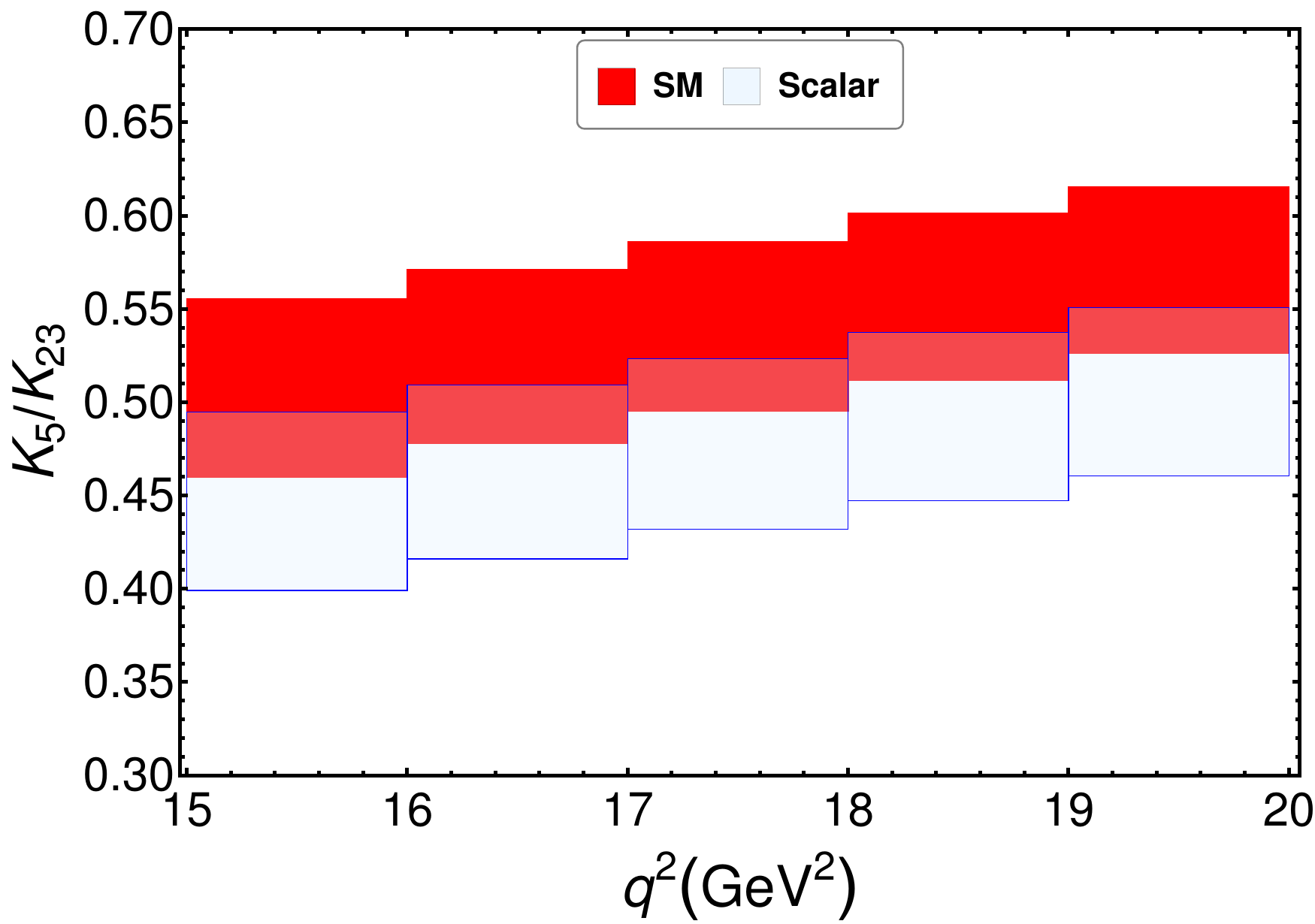}
		\includegraphics[scale=0.4]{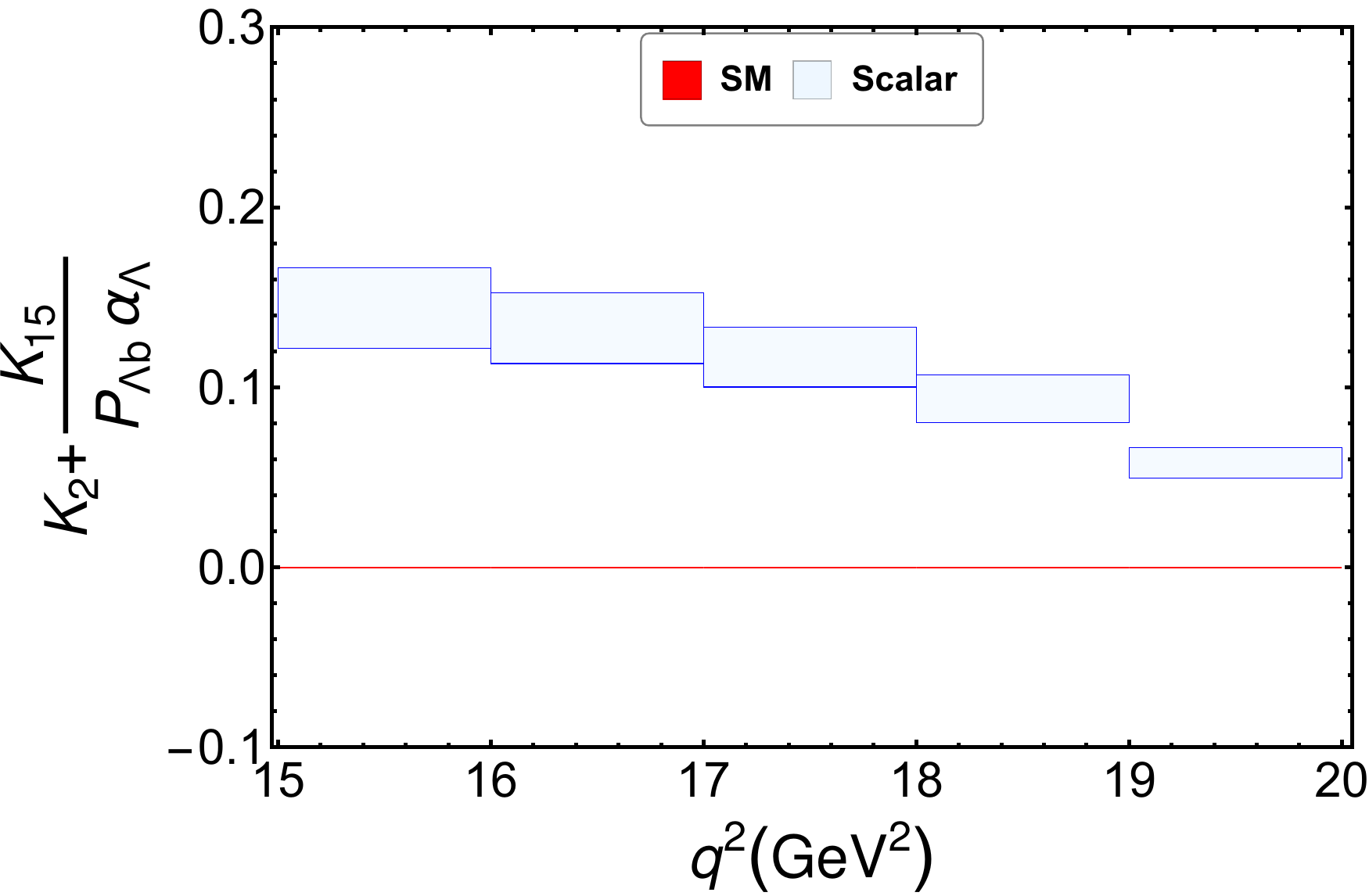}
		\caption{The angular observables $K_5/K_{23}$ (left) and $K_2 + K_{15}/(P_{\Lambda_b} \alpha_\Lambda)$ (right) for polarized $\Lambda_b\to\Lambda(\to p\pi)\mu^+\mu^-$ decay in different $q^2$ bins in the SM and in the presence of SP operators. The set of scalar couplings are $\mC_S = -1.5 + i  0.03, \mC_S^{\prime} = 2.2 + i 0.2, \mC_P = 2.5 + i 0.1, \mC_P^{\prime} = -3-i 0.04 $.
			\label{fig:NPplots2}}
	\end{center} 
\end{figure}

\begin{table}[!htb]
	\caption{\small Expected experimental precision on the angular observables achievable at the future LHCb. The second column shows the precision achieved at the LHCb using combined Run 1 and Run 2 with ~650 events in the $15<q^2<20$GeV$^2$ bin \cite{Aaij:2018gwm}. At future LHCb, about $\sim 8000$ events are expected with 50fb$^{-1}$ integrated luminosity and about 50000 events are expected with integrated luminosity of $300$fb$^{-1}$ \cite{Blake:2017une}. Precision achievable at these luminosities are shown in the third and the fourth columns. \label{tab:1}}
	\centering
	\renewcommand{\arraystretch}{1}
	\begin{tabular}{lccc|lccc}
		\hline
		Obs. & Run\,1 + Run\,2   &  $50$fb$^{-1}$  &$300$fb$^{-1}$  & Obs. &  Run\,1 + Run\,2 &  $50$fb$^{-1}$  & $300$fb$^{-1}$  \\
		\hline
		$M_{1}$   & 0.020 &0.006&0.002 &$M_{18}$ & 0.058 &0.017&0.007 \\
		$M_{2}$   & 0.040 &0.011&0.005 &$M_{19}$& 0.122 &0.035&0.014\\
		$M_{3}$  &0.029 &0.008& 0.0033 & $M_{20}$ &0.056 &0.016& 0.006\\
		$M_{4}$ & 0.046 &0.013& 0.0052& $M_{21}$ &0.105  &0.030&0.012\\
		$M_{5}$ &0.081 &0.023& 0.009 &$M_{22}$  & 0.045  &0.013&0.005\\
		$M_{6}$ & 0.055 &0.016&0.006& $M_{23}$  & 0.077  &0.022&0.009\\
		$M_{7}$ & 0.084 &0.024&0.010& $M_{24}$ &0.033 &0.009&0.004 \\
		$M_{8}$ &0.037  &0.011&0.004 &$M_{25}$ & 0.176  &0.050&0.020\\
		$M_{9}$  & 0.084 &0.024&0.009 &$M_{26}$ &0.074  &0.021&0.008\\
		$M_{10}$  &0.037 &0.011& 0.004&$M_{27}$ &  0.140  &0.040&0.016\\
		$M_{11}$  &  0.043 &0.012&0.005 &$M_{28}$ & 0.058&0.017& 0.007\\
		$M_{12}$  & 0.063 &0.018& 0.007&$M_{29}$ & 0.097 &0.028& 0.011\\
		$M_{13}$  & 0.045 &0.013&0.005 &$M_{30}$& 0.061 &0.017& 0.007 \\
		$M_{14}$  &  0.082 &0.023&0.009 &$M_{31}$&0.094 &0.027& 0.011 \\
		$M_{15}$  &0.117 &0.033&0.013 &$M_{32}$  & 0.055 &0.016&0.006 \\
		$M_{16}$  &0.084 &0.024&0.001& $M_{33}$&  0.060  &0.017&0.006 \\
		$M_{17}$  & 0.120  &0.034&0.014 &$M_{34}$& 0.058 &0.017&0.006\\
		\hline
	\end{tabular}
\end{table}

The relations \eqref{eq:K57relation} and \eqref{eq:K523relation} are quite interesting. In the SM+SM$^\prime$ they are independent of any short-distance physics but are modified by SP operators. These are null test of the SM+SM$^\prime$. In figure \ref{fig:NPplots2} we show the ratio $K_5/K_{23}$ in the SM and for set of representative scalar couplings in different $q^2$ bins. The relations \eqref{combo:rhoS1} and \eqref{eq:K215relation} are the ones where the combinations of angular observables depend on the SP operators only. In \ref{fig:NPplots2} figure we plot $K_2 + K_{15}/(P_{\Lambda_b} \alpha_\Lambda)$ for a set of scalar couplings. This combination is also a null test of SM+SM$^\prime$.

In ref.~\cite{Aaij:2018gwm} the observables have been measured in the $15<q^2<20$GeV$^2$ bin using combined Run 1 and Run 2 data sets that correspond to $610\pm 29$ events. Due to large uncertainties, the polarization observables are consistent with zero. This is expected as the $P_{\Lambda_b}$ observed in \cite{Aaij:2013oxa,CMS:2016iaf} is also consistent with zero. In table \ref{tab:1} we show the precision achievable at future LHCb luminosities. Effects of backgrounds are expected to have a negligible effect on experimental precision and are neglected. As can be seen from the table, a large number of events are expected at future LHCb luminosities, and the precision on angular observables is expected to improve. The precision with which an observable can be measured is independent of the $P_{\Lambda_b}$. But as the polarized observables are proportional to $P_{\Lb}$, it affects the sensitivity with which NP can be disentangled from the SM, with lower $P_{\Lb}$ leading to a lower sensitivity. This caveat is in order when comparing the table \ref{tab:1} with the plots \ref{fig:NPplots1} and \ref{fig:NPplots2} which are obtained with $P_{\Lambda_b}=1$. Given that a 10\% polarization of $\Lambda_b$ cannot be excluded, the tests discussed in section \ref{sec:lowrecoil} remain one of the interesting possibilities in the polarized $\Lambda_b\to\Lambda(\to p\pi)\ell^+\ell^-$ decay.

\section{Summary \label{sec:summary}}
In this paper we derive a full angular distribution of polarized $\Lambda_b$ decay $\Lambda_b\to\Lambda(\to p\pi)\ell^+\ell^-$. The distribution is obtained for a set of operators where the Standard Model operator basis is supplemented with its chirality flipped counterparts and additional scalar and pseudoscalar operators. We retain the mass of the final state leptons. We apply a Heavy Quark Effective Theory framework valid at low hadronic recoil and obtain factorization of long and short-distance physics in the angular observables. Using the factorized expressions we construct several tests of form factors and Wilson coefficients, including some null test of the Standard Model and its chirality flipped counterparts. Our analysis shows that new insight to $b\to s\ell^+\ell^-$ transition can be obtained from this mode.

\section*{Acknowledgements}
We would like to thank Vishal Bhardwaj, Luis Miguel Garcia,  and Tanumoy Mandal for useful discussions. DD acknowledges the DST, Govt. of India for the INSPIRE Faculty Fellowship (grant no. IFA16-PH170).  

\appendix

\section{Nucleon spinor in $\Lambda$ rest frame \label{sec:LamRF}}
The angles $\theta_b$ and $\phi_b$ made by the $p$ in the $p\pi$ rest frame (denoted as $p\pi$-RF) are shown in figure \ref{Fig:angular}. In this frame, characterized by $k^2=m_\Lam^2$, the four-momentum $k^\mu_{1,2}$ read 
\begin{equation}
\begin{split}
&k_1^\mu\big|_{p\pi-\rm RF} = (E_p, |k_{p\pi}|\sin\theta_b\cos\phi_b, |k_{p\pi}|\sin\theta_b\sin\phi_b, |k_{p\pi}|\cos\theta_b )\, ,\\
&k_2^\mu\big|_{p\pi-\rm RF} = (E_\pi, -|k_{p\pi}|\sin\theta_b\cos\phi_b, -|k_{p\pi}|\sin\theta_b\sin\phi_b, -|k_{p\pi}|\cos\theta_b )\, ,
\end{split}
\end{equation}
where
\begin{equation}
E_p = \frac{k^2+m_p^2-m_{\pi}^2}{2\sqrt{k^2}}\, ,\quad E_\pi = \frac{k^2+m_{\pi}^2-m_p^2}{2\sqrt{k^2}}\, ,
\end{equation}
and 
\begin{equation}
|k_{p\pi}| = \frac{\sqrt{\lambda(k^2, m_p^2, m_{\pi}^2)}}{2\sqrt{k^2}}\, .
\end{equation}
The Dirac spinors for the $\Lambda$ are
\begin{equation}
u(k,+1/2)=\left(\begin{array}{c} 
\sqrt{2m_\Lambda}\\ 
0\\ 
0\\
0\end{array}\right), \qquad
u(k,-1/2)=
\left(\begin{array}{c} 
0\\ 
\sqrt{2m_\Lambda}\\ 
0\\
0\end{array}\right), \qquad
\end{equation}
and the spinor for $p$ are \cite{Haber:1994pe}
\begin{equation}
u(k_1,+1/2)=\frac{1}{\sqrt{2m_\Lam}}\left(\begin{array}{c} 
\sqrt{r_+}\cos\frac{\theta_b}{2}\\ 
\sqrt{r_+}\sin\frac{\theta_b}{2}e^{i\phi_b}\\ 
\sqrt{r_-}\cos\frac{\theta_b}{2}\\
\sqrt{r_-}\sin\frac{\theta_b}{2}e^{i\phi_b} \end{array}\right), \qquad
u(k_1,-1/2)=\frac{1}{\sqrt{2m_\Lam}}\left(\begin{array}{c} 
-\sqrt{r_+}\sin\frac{\theta_b}{2}e^{-i\phi_b}\\ 
\sqrt{r_+}\cos\frac{\theta_b}{2}\\ 
\sqrt{r_-}\sin\frac{\theta_b}{2}e^{-i\phi_b} \\
-\sqrt{r_-}\cos\frac{\theta_b}{2}  \end{array}\right) .
\end{equation}
The secondary weak decay is governed by the Hamiltonian  
\begin{equation}
H^{\rm eff}_{\Delta S=1} = \frac{4G_F}{\sqrt{2}} V_{ud}^\ast V_{us} [\bar{d} \gamma_\mu P_L u] [\bar{u} \gamma^\mu P_L s]\, .
\end{equation}
The $\Lambda(k,s_\Lambda)\to p(k_1,s_p)\pi(k_2)$ matrix elements are parametrized as
\begin{align}
H_2(s_\Lambda, s_p) & \equiv
\bra{p(k_1, s_p) \pi^-(k_2)} \left[\bar{d} \gamma_\mu P_L u\right]\left[\bar{u} \gamma^\mu P_L s\right]
\ket{\Lambda(k, s_\Lambda)} \cr
& = \big[\bar u(k_1, s_p) \big(\xi \,\gamma_5 + \omega\big) u(k, s_\Lambda)\big] \,.
\end{align}
In terms of the kinematics variables the secondary decay amplitudes are
\begin{align}
H_2(+1/2,+1/2) &= \left(\sqrt{r_+} \, \omega - \sqrt{r_-} \, \xi \right) \cos\frac{\theta_\Lambda}{2} \,, \cr
H_2(+1/2,-1/2) &= -\left(\sqrt{r_+} \, \omega + \sqrt{r_-} \, \xi \right) \sin\frac{\theta_\Lambda}{2}\,
e^{i \phi}  \,, \cr
H_2(-1/2,+1/2) &= -\left(-\sqrt{r_+} \, \omega + \sqrt{r_-} \, \xi \right) \sin\frac{\theta_\Lambda}{2}\,
e^{-i \phi} \,, \cr
H_2(-1/2,-1/2) &= \left(\sqrt{r_+}\, \omega + \sqrt{r_-} \,\xi \right) \cos\frac{\theta_\Lambda}{2} \,.
\end{align}
In the final distribution only the parity violating parameter is relevant
\begin{equation}
\alpha = \frac{-2\re(\omega\xi)}{ \sqrt{\frac{r_-}{r_+}} |\xi|^2 + \sqrt{\frac{r_+}{r_-}} |\omega|^2  } \equiv \alpha^{\rm exp} \, .
\end{equation}

\section{Leptonic helicity amplitudes \label{sec:LepHel}}
To calculate the lepton helicity amplitudes for $\Lambda_b(p,s_p)\to \Lambda(k,s_k)$ $\ell^+(q_+)\ell^-(q_-)$ in the dilepton rest-frame we have to calculate the following two currents
\begin{equation}
\bar{u}_{\ell_1}(1\mp\gamma_5)v_{\ell_2}\, ,\quad\text{and}\quad \bar{\epsilon}^\mu(\lambda)\bar{u}_{\ell_1}\gamma_\mu(1\mp\gamma_5)v_{\ell_2}\,.
\end{equation}
In the dilepton rest-frame ($2\ell\rm RF$), the four momentum of the two leptons are
\begin{align}
& q_-^\mu \Big|_{2\ell\rm RF} = (E_\ell, -|q_{2\ell}|\sin\theta_\ell\cos\phi_\ell, -|q_{2\ell}|\sin\theta_\ell\sin\phi_\ell, -|q_{2\ell}|\cos\theta_\ell)\, ,\\
&q_+^\mu \Big|_{2\ell\rm RF} = (E_\ell, |q_{2\ell}|\sin\theta_\ell\cos\phi_\ell, |q_{2\ell}|\sin\theta_\ell\sin\phi_\ell, |q_{2\ell}|\cos\theta_\ell)\, ,
\end{align}
with 
\begin{equation}
|q_{2\ell}| = \frac{\beta_\ell}{2}\sqrt{q^2}\, ,\quad\quad E_\ell = \frac{\sqrt{q^2}}{2}\, ,\quad\quad \beta_\ell = \sqrt{1-\frac{4m_\ell^2}{q^2}}\, .
\end{equation}
Following \cite{Haber:1994pe} we give the explicit expressions of the spinors in this frame are
\begin{align}
& u_{\ell^-}(\lambda) = 
\begin{pmatrix}
\sqrt{E_\ell+m_\ell} \chi^u_\lambda  \\ 2 \lambda \sqrt{E_\ell-m_\ell} \chi^u_\lambda
\end{pmatrix}\, ,
\quad \chi^u_{+\frac{1}{2}} = \begin{pmatrix} \cos\frac{\theta_\ell}{2} \\ -e^{-i\phi_l}\sin\frac{\theta_\ell}{2} \end{pmatrix}\, ,
\quad \chi^u_{-\frac{1}{2}} = \begin{pmatrix} e^{i\phi_l}\sin\frac{\theta_\ell}{2} \\ \cos\frac{\theta_\ell}{2} \end{pmatrix}\, ,\\
& v_{\ell^+}(\lambda) = 
\begin{pmatrix}
\sqrt{E_\ell-m_\ell} \chi^v_{-\lambda}  \\ -2 \lambda \sqrt{E_\ell+m_\ell} \chi^v_{-\lambda}
\end{pmatrix}\, ,
\quad \chi^v_{+\frac{1}{2}} = \begin{pmatrix} \sin\frac{\theta_\ell}{2} \\ e^{-i\phi_l}\cos\frac{\theta_\ell}{2} \end{pmatrix}\, ,
\quad \chi^v_{-\frac{1}{2}} = \begin{pmatrix} -e^{i\phi_l}\cos\frac{\theta_\ell}{2} \\ \sin\frac{\theta_\ell}{2} \end{pmatrix}\, .
\end{align}
Using these spinors we obtain the following nonzero expressions of the leptonic helicity amplitudes for different combinations of $\lambda_1$ and $\lambda_2$ 
\begin{align}
& L^{\plpl}_L =-\sqrt{q^2}(1+\beta_\ell)e^{i\phi_l}\, ,
\quad L^{\mimi}_R = \sqrt{q^2}(1+\beta_\ell)e^{-i\phi_l}\, ,\\ 
& L^{\mimi}_L =-\sqrt{q^2}(1-\beta_\ell)e^{-i\phi_l}\, ,
\quad L^{\plpl}_R = \sqrt{q^2}(1-\beta_\ell)e^{i\phi_l}\, ,\\
& L^{\plpl}_{L,+1}=  \sqrt{2}m_\ell\sin\theta_\ell e^{2 i\phi_l}\, ,
\quad L^{\plpl}_{R,+1}=  \sqrt{2}m_\ell\sin\theta_\ell e^{2 i\phi_l}\, ,\\
&
L^{\mimi}_{L,-1}= \sqrt{2}m_\ell\sin\theta_\ell e^{-2 i\phi_l}\, ,
\quad L^{\mimi}_{R,-1} = \sqrt{2}m_\ell\sin\theta_\ell e^{-2 i\phi_l} \, ,\\
& L^{\mimi}_{L,+1}=-\sqrt{2}m_\ell\sin\theta_\ell\, ,
\quad L^{\mimi}_{R,+1}=-\sqrt{2}m_\ell\sin\theta_\ell\, ,\\
&
L^{\plpl}_{L,-1}=-\sqrt{2}m_\ell\sin\theta_\ell\, ,
\quad L^{\plpl}_{R,-1} = -\sqrt{2}m_\ell\sin\theta_\ell\, ,\\
& L^{\plmi}_{L,+1} =\sqrt{\frac{q^2}{2}}(1-\beta_\ell)(1-\cos\theta_\ell)e^{i\phi_l}\, ,
\quad L^{\mipl}_{R,-1} = -\sqrt{\frac{q^2}{2}}(1-\beta_\ell)(1-\cos\theta_\ell)e^{-i\phi_l}\, ,\\
& L^{\mipl}_{L,+1} =- \sqrt{\frac{q^2}{2}}(1+\beta_\ell)(1+\cos\theta_\ell)e^{i\phi_l}\, ,
-L^{\plmi}_{R,-1} = -\sqrt{\frac{q^2}{2}}(1+\beta_\ell)(1+\cos\theta_\ell)e^{-i\phi_l}\, ,\\
& L^{\plmi}_{R,+1}=\sqrt{\frac{q^2}{2}}(1+\beta_\ell)(1-\cos\theta_\ell)e^{i\phi_l}\, ,
L^{\mipl}_{L,-1} =- \sqrt{\frac{q^2}{2}}(1+\beta_\ell)(1-\cos\theta_\ell)e^{-i\phi_l}\, ,\\
& L^{\mipl}_{R,+1}= -\sqrt{\frac{q^2}{2}}(1-\beta_\ell)(1+\cos\theta_\ell)e^{i\phi_l} \, ,
-L^{\plmi}_{L,-1} = -\sqrt{\frac{q^2}{2}}(1-\beta_\ell)(1+\cos\theta_\ell)e^{-i\phi_l}\, ,\\
%
%
& L^{\plpl}_{L,0}= 2m_\ell\cos\theta_\ell e^{i\phi_l}\, ,
L^{\mimi}_{L,0}= -2m_\ell\cos\theta_\ell e^{-i\phi_l}\, ,\\
&
L^{\plpl}_{R,0}= 2m_\ell\cos\theta_\ell e^{i\phi_l} \, ,
L^{\mimi}_{R,0} = -2m_\ell\cos\theta_\ell e^{-i\phi_l}\, ,\\
& L^{\plmi}_{L,0}=  \sqrt{q^2}(1-\beta_\ell)\sin\theta_\ell  \, ,
L^{\mipl}_{R,0} = \sqrt{q^2}(1-\beta_\ell)\sin\theta_\ell\, ,\\
& L^{\mipl}_{L,0} =  \sqrt{q^2}(1+\beta_\ell)\sin\theta_\ell \, ,
L^{\plmi}_{R,0} = \sqrt{q^2}(1+\beta_\ell)\sin\theta_\ell\, ,\\
& L^{\plpl}_{L,t} =-2m_\ell  e^{i\phi_l}, 
L^{\mimi}_{L,t} = -2m_\ell e^{-i\phi_l}\, ,\\
& L^{\plpl}_{R,t} =2m_\ell e^{i\phi_l}, 
L^{\mimi}_{R,t} = 2m_\ell e^{-i\phi_l}\, .
\end{align}
The combinations of $\lambda_1$ and $\lambda_2$ for which the amplitudes vanish are not shown.

\section{Angular coefficients \label{sec:Ks}}
The expressions of angular observables $K_i$ are given below. With reference to equation \eqref{eq:Ki} we separately show the massless part $\mathcal{K}_i$ and the massive parts $\mathcal{K}^{\prime}_i$, $\mathcal{K}^{\prime\prime}_i$. The expressions of $K_i$ extend previous expressions for the SM+SM$^\prime$ amplitudes \cite{Blake:2017une} by including SP amplitudes. The expressions of $\mathcal{K}^{\prime}_i$, $\mathcal{K}^{\prime\prime}_i$ given below are new in this paper.
\begin{eqnarray}\label{eq:KKprime1}
\mK_{1} &=& \frac{1}{4} \bigg( 2|\ARpa0|^2 + |\ARpa1|^2 + 2|\ARpe0|^2 + |\ARpe1|^2 + \{ R \leftrightarrow L  \} \bigg) \nn\\&
+& \frac{1}{2}\bigg( |\ARSPp|^2 +| \ARSPm|^2+ \{ R \leftrightarrow L  \}  \bigg)\, ,~~~~\\
\mK_{1}^\prime &=& \re\bigg( \ARSPp A_{\perp t}^\ast+   \ARSPm A_{\|t}^\ast  -     \{ R \leftrightarrow L  \} \bigg)\, , \\
\mK_{1}^{\prime\prime} &=& -\bigg( |A^R_{\|_0}|^2 + |A^R_{\perp_0}|^2 + \{ R \leftrightarrow L \} \bigg) + \bigg( |A_{\perp t}|^2   + \{ \perp \leftrightarrow \| \}\bigg)  \nn\\ &
+& 2\re\bigg( A^R_{\perp_0}A^{\ast L}_{\perp_0} + A^R_{\perp_1}A^{\ast L}_{\perp_1} + \{ \perp \leftrightarrow \| \}  \bigg)\nn\\ &
-&  \bigg( |\ARSPp|^2 +| \ARSPm|^2+ \{ R \leftrightarrow L  \}  \bigg)
-\re \bigg( \ARSPm \AsLSPm + \ARSPp \AsLSPp   \bigg)
\, , \\
%
\mK_{2} &=& \frac{1}{2}\bigg( |\ARpa1|^2 + |\ARpe1|^2 + \{R \leftrightarrow L \} \bigg) 
+ \frac{1}{2}\bigg( |\ARSPp|^2 +| \ARSPm|^2+ \{ R \leftrightarrow L  \}                       \bigg) \, ,\\
\mK_{2}^\prime &=& \re\bigg( A_{\perp t}\AsRSPp+ A_{\| t} \AsRSPm- \{ R \leftrightarrow L  \}   \bigg) ,\\
\mK_{2}^{\prime\prime} &=& \bigg( |A^R_{\|_0}|^2 - |A^R_{\|_1}|^2 + |A^R_{\perp_0}|^2 - |A^R_{\perp_1}|^2 + \{ R \leftrightarrow L \} \bigg) + \bigg( |A_{\perp t}|^2  + \{ \perp \leftrightarrow \| \}\bigg)\, \nn\\ 
&-&\bigg( |\ARSPp|^2 +| \ARSPm|^2+ \{ R \leftrightarrow L  \}   \bigg)\, \nn\\ 
&+& 2\re\bigg( A^R_{\perp_0}A^{\ast L}_{\perp_0} + A^R_{\perp_1}A^{\ast L}_{\perp_1} -\ARSPp \AsLSPp+ \{\perp \leftrightarrow \| \} \bigg)\, , \\
\mK_{3} &=& -\beta_\ell \bigg( A^R_{\perp_1}A^{\ast R}_{\|_1} - \{ R \leftrightarrow L \}  \bigg)\, ,\\
\mK_{3}^\prime &=& \beta_\ell \re\bigg(
\ARSPm \AsRpa0 + \ARSPp \AsRpe0 
+ \{ R \leftrightarrow L \} \nn\\
&+& \ALSPm \AsRpa0 + \ARSPm \AsLpa0 +  \{  \| \leftrightarrow  \perp \}\bigg)\, ,\\
\mK_{3}^{\prime\prime} &=& 0\, ,\\
%
\mK_{4} &=& \frac{\alpha_\Lambda}{2}
\re\bigg( 2 A^R_{\perp_0}A^{\ast R}_{\|_0} + A^R_{\perp_1}A^{\ast R}_{\|_1} +2 \ARSPp \AsRSPm+ \{ R \leftrightarrow L \} \bigg)  \, ,\\
\mK_{4}^\prime &=& \alpha_\Lambda \re\bigg( \ARSPm A_{\perp t}^{\ast} + \ARSPp A_{\| t}^{\ast}- \{ R \leftrightarrow L \}  \bigg)\, ,\\
\mK_{4}^{\prime\prime} &=& -2 \alpha_\Lambda \re \bigg(
A^R_{\perp_0}A^{\ast R}_{\|_0} + \ARSPp \AsRSPm +\{ R \leftrightarrow L \}\nn\\
&-& A^R_{\perp_0}A^{\ast L}_{\|_0} - A^R_{\perp_1}A^{\ast L}_{\|_1} + \{ \| \leftrightarrow \perp \} \nn\\
&-& A_{\perp t}A^\ast_{\|t} + \ALSPp \AsRSPm +\ARSPp \AsLSPm \bigg)\, ,\\
\mK_{5} &=& \alpha_\Lambda \re \bigg( A^R_{\perp_1}A^{\ast R}_{\|_1} 
+ \ARSPp \AsRSPm +  \{ R \leftrightarrow L \}  \bigg) \, ,\\
\mK_{5}^\prime &=& \alpha_\Lambda \re\bigg( \ARSPm A_{\perp t}^{\ast} + \ARSPp A_{\| t}^{\ast} +  \{ R \leftrightarrow L \} \bigg)\, ,\\
\mK_{5}^{\prime\prime} &=& 2\alpha_\Lambda \re\bigg(
A^R_{\perp_0}A^{\ast R}_{\|_0}-A^R_{\perp_1}A^{\ast R}_{\|_1} + \{ R \leftrightarrow L \}\nn\\ 
&+&A^R_{\perp_0}A^{\ast L}_{\|_0} + A^R_{\|_0}A^{\ast L}_{\perp_0}
+ A^R_{\perp_1}A^{\ast L}_{\|_1} +A^R_{\|_1}A^{\ast L}_{\perp_1} \nn\\
&+& A_{\perp t}A^{\ast}_{\|t}
-(\ARSPp \AsRSPm + \{ R \leftrightarrow L \} +\ALSPp \AsRSPm +\ARSPp \AsLSPm)
\bigg)\, ,\\
\mK_{6} &=& -\frac{\alpha_\Lambda\beta_\ell}{2} \bigg( |\ARpe1|^2 + |\ARpa1|^2 - \{ R \leftrightarrow L \}   \bigg)\, ,\\
\mK_{6}^\prime &=& \alpha_\Lambda\beta_\ell 
\re\bigg( \ARSPp \AsRpa0 +\ARSPm \AsRpe0 + \{ R \leftrightarrow L \} \nn\\
&+&\ARSPm \AsLpe0 + \ALSPm \AsRpe0 + \{ \| \leftrightarrow \perp \} 
\bigg)\, ,\\
\mK_{6}^{\prime\prime} &=& 0\, ,\\
%
\mK_{7} &=& \frac{\alpha_\Lambda}{\sqrt{2}} \re\bigg( A^R_{\perp_1}A^{\ast R}_{\|_0} - A^R_{\|_1}A^{\ast R}_{\perp_0} + \{ R \leftrightarrow L \} \bigg)\, ,\\
\mK_{7}^\prime &=& 0\, ,\\
\mK_{7}^{\prime\prime} &=& 2\sqrt{2}\alpha_\Lambda \re\bigg( A^R_{\|_1}A^{\ast R}_{\perp_0} - A^R_{\perp_1}A^{\ast R}_{\|_0} + \{ R \leftrightarrow L \}  \bigg)\, ,\\
\mK_{8} &=& \frac{\alpha_\Lambda\beta_\ell}{\sqrt{2}} \re \bigg( A^R_{\perp_1}A^{\ast R}_{\perp_0} - A^R_{\|_1}A^{\ast R}_{\|_0} - \{ R \leftrightarrow L \} \bigg)\, ,\\
\mK_{8}^\prime &=& \frac{\alpha_\Lambda\beta_\ell}{\sqrt{2}} \re \bigg( 
\ARSPp \AsRpa1 -\ARSPm \AsRpe1 + \{ R \leftrightarrow L \} \nn\\
&+& \ALSPp \AsRpa1 + \ARSPp \AsLpa1 -  \{ \| \leftrightarrow \perp \} \bigg)\, ,\\
\mK_{8}^{\prime\prime}  &=& 0\, ,\\
\mK_{9} &=& \frac{\alpha_\Lambda}{\sqrt{2}} \im \bigg( \ARpe1 \AsRpe0 - \ARpa1 \AsRpa0 + \{ R \leftrightarrow L \} \bigg)\, ,\\
\mK_{9}^\prime &=& 0\, ,\\
\mK_{9}^{\prime\prime} &=& 2\sqrt{2}\alpha_\Lambda \im \bigg( A^R_{\|_1}A^{\ast R}_{\|_0} - A^R_{\perp_1}A^{\ast R}_{\perp_0} + \{ R \leftrightarrow L \}  \bigg)\, ,\\
%
\mK_{10} &=& \frac{\alpha_\Lambda\beta_\ell}{\sqrt{2}} \im \bigg( A^R_{\perp_1}A^{\ast R}_{\|_0} - A^R_{\|_1}A^{\ast R}_{\perp_0} - \{ R \leftrightarrow L \} \bigg)\, ,\\
\mK_{10}^\prime &=& \frac{\alpha_\Lambda\beta_\ell}{\sqrt{2}}
\im\bigg( \ARSPp \AsRpe1 - \ARSPm \AsRpa1 + \{ R \leftrightarrow L \} \nn\\
&+&   \ALSPp \AsRpe1 +\ARSPp \AsLpe1 + \{ \perp \leftrightarrow \| \} \bigg)\, ,\\
\mK_{10}^{\prime\prime} &=& 0\, ,\\
%
\mK_{11} &=& \frac{P_{\Lambda_b}}{2}\re\bigg( 2 \ARpa0 \AsRpe0 - \ARpa1 \AsRpe1 + \{ R \leftrightarrow L \}      \nn\\
&+& 2 \ARSPp \AsRSPm + \{ R \leftrightarrow L \}  \bigg) \, ,\\ 
\mK_{11}^\prime &=& P_{\Lambda_b} \re \bigg( \ARSPp A_{\|t}^* + \ARSPm  A_{\perp t}^* - \{ R \leftrightarrow L \}   \bigg)
\, ,\\ 
\mK_{11}^{\prime\prime} &=& -2 P_{\Lambda_b} \re  \bigg(
\ARpa0 \AsRpe0 + \ARSPp \AsRSPm + \{ R \leftrightarrow L \} -A_{\perp t} A_{\|t}^* \nn\\
&+&( \ARpa1 \AsLpe1 -\ARpa0 \AsLpe0  + \ALSPp \AsRSPm + \{ \| \leftrightarrow \perp \} )
\bigg)\, ,\\
\mK_{12} &=& -P_{\Lambda_b} \re \bigg( \ARpa1 \AsRpe1 + \{ R \leftrightarrow L \} - \ARSPp \AsRSPm - \{ R \leftrightarrow L \} \bigg)   \, ,\\ 
\mK_{12}^\prime &=& P_{\Lambda_b} \re \bigg( \ARSPp A_{\| t}^* +\ARSPm A_{\perp t} - \{ R \leftrightarrow L \} \bigg) \, ,\\ 
\mK_{12}^{\prime\prime} &=& 2 P_{\Lambda_b} \re \bigg( 
\ARpa0 \AsRpe0 + \ARpa1 \AsRpe1 -\ARSPp \AsRSPm +\{ R \leftrightarrow L \}\nn\\
&+& A_{\perp t} A_{\|}^*\nn\\
&+& \ARpa0 \AsLpe0 - \ARpa1  \AsLpe1 +\{ \| \leftrightarrow \perp \}\nn\\
& -& \ALSPp \AsRSPm -\{ R \leftrightarrow L \} \bigg)  \, ,\\
\mK_{13} &=& \frac{P_{\Lambda_b} \beta_l}{2} \bigg(
|\ARpa1|^2 + |\ARpe1|^2  - \{ R \leftrightarrow L \}  \bigg)   \, ,\\ 
\mK_{13}^\prime &=& P_{\Lambda_b} \beta_l \re \bigg( 
\ARSPm \AsRpe0 + \ARSPp \AsRpa0 + \{ R \leftrightarrow L \}\nn\\
&+& \ARSPm \AsLpe0 + \ALSPp \ARpe0 +\{ \| \leftrightarrow \perp \}
\bigg) \, ,\\ 
\mK_{13}^{\prime\prime} &=& 0 \, ,\\
\mK_{14} &=& \frac{P_{\Lambda_b} \alpha_\Lambda}{4}  \re \bigg( 
2 |\ARpa0|^2 +2 |\ARpe0|^2 -|\ARpa1|^2- |\ARpe1|^2 +\{ R \leftrightarrow L \} \nn\\
&+& 2(|\ARSPm|^2+|\ARSPp|^2 +\{ R \leftrightarrow L \} ) \bigg)   \, ,\\ 
\mK_{14}^\prime &=& P_{\Lambda_b} \alpha_\Lambda \re \bigg( 
\ARSPp A_{\perp t}^* + \ARSPm A_{\| t}^* - \{ R \leftrightarrow L \} 
\bigg) \, ,\\ 
\mK_{14}^{\prime\prime} &=& -P_{\Lambda_b} \alpha_\Lambda \bigg(   
|\ARpa0|^2 +|\ARpe0|^2 + \{ R \leftrightarrow L \} \nn\\
&-& ( |A_{\|t} |^2 +|A_{\perp t} |^2)\nn\\
&+& (|\ARSPp|^2+ |\ARSPm|^2 +\{ R \leftrightarrow L \} )\nn\\
&-&2 \re (\ALpa0 \AsRpa0 - \ALpa1 \AsRpa1  -\ARSPm \AsLSPm+\{ \| \leftrightarrow \perp \} )\bigg)  \, ,\\
%
\mK_{15} &=&\frac{ P_{\Lambda_b} \alpha_\Lambda}{2} \bigg(   
-( |\ARpe1|^2 + |\ARpa1|^2 +\{ R \leftrightarrow L \}) \nn\\
&+& (|\ARSPm |^2 + |\ARSPp |^2 +\{ R \leftrightarrow L \} ) \bigg)   \, ,\\ 
\mK_{15}^\prime &=& P_{\Lambda_b} \alpha_\Lambda \re \bigg( \ARSPm  A_{\|t}^* + \ARSPp A_{\perp t} ^* -\{ R \leftrightarrow L \}   \bigg) \, ,\\ 
\mK_{15}^{\prime\prime} &=& P_{\Lambda_b} \alpha_\Lambda
\bigg[ |\ARpa0|^2 + |\ARpa1|^2 + |\ARpe0|^2 + |\ARpe1|^2 +\{ R\leftrightarrow L \} \nn\\
&+& (|A_{\|t} |^2 + |A_{\perp t} |^2 )-(|\ARSPp|^2+| \ARSPm |^2 +\{ R\leftrightarrow L \} )\nn\\
& +& 2\re (\ALpa0 \AsRpa0- \ALpa1 \AsRpa1 -\ARSPm \AsLSPm
+\{ \| \leftrightarrow \perp \} ) \bigg]  \, ,\\
\mK_{16} &=& P_{\Lambda_b} \alpha_\Lambda \beta_l   \re \bigg(
\ARpa1 \AsRpe1 - \{ R \leftrightarrow L \}  \bigg)   \, ,\\ 
\mK_{16}^\prime &=& P_{\Lambda_b} \alpha_\Lambda \beta_l  \re  \bigg( 
\ARSPm \AsRpa0 + \ARSPp \AsRpe0 +\{ R \leftrightarrow L \}\nn\\
&+& \ARSPm \AsLpa0 + \ALSPm \AsRpa0 + \{ \| \leftrightarrow \perp \}
\bigg) \, ,\\ 
\mK_{16}^{\prime\prime} &=& 0 \, ,\\
\mK_{17} &=& -\frac{P_{\Lambda_b} \alpha_\Lambda}{\sqrt{2}}  \re \bigg(
\ARpa1 \AsRpa0 - \ARpe1 \AsRpe0 +  \{ R \leftrightarrow L \}  \bigg)   \, ,\\ 
\mK_{17}^\prime &=& 0\, ,\\ 
\mK_{17}^{\prime\prime} &=&2\sqrt{2}P_{\Lambda_b} \alpha_\Lambda \re \bigg( \ARpa1 \AsRpa0 - \ARpe1 \AsRpe0 +  \{ R \leftrightarrow L \}   \bigg)  \, ,\\
%
\mK_{18} &=& \frac{P_{\Lambda_b}\alpha_\Lambda \beta_l}{\sqrt{2}}  \re \bigg(  \ARpe1 \AsRpa0 - \ARpa1 \AsRpe0 -
\{ R \leftrightarrow L \}  \bigg)   \, ,\\ 
\mK_{18}^\prime &=& \frac{P_{\Lambda_b}\alpha_\Lambda \beta_l}{\sqrt{2}}  \re \bigg(   \ARSPm \AsRpa1 - \ARSPp \AsRpe1 +\{ R \leftrightarrow L \} \nn\\
&+& \ALSPm \AsRpa1 +\ARSPm \AsLpa1 -\{ \| \leftrightarrow \perp \}
\bigg) \, ,\\ 
\mK_{18}^{\prime\prime} &=&0  \, ,\\
%
\mK_{19} &=& -\frac{P_{\Lambda_b}\alpha_\Lambda }{\sqrt{2}} 
\im \bigg(  \ARpa1 \AsRpe0 - \ARpe1 \AsRpa0 - \{ R \leftrightarrow L \}  \bigg)   \, ,\\ 
\mK_{19}^\prime &=&0 \, ,\\ 
\mK_{19}^{\prime\prime} &=& 2 \sqrt{2} P_{\Lambda_b}\alpha_\Lambda \im \bigg( \ARpa1 \AsRpe0 - \ARpe1 \AsRpa0 - \{ R \leftrightarrow L \}    \bigg)  \, ,\\
\mK_{20} &=& \frac{P_{\Lambda_b}\alpha_\Lambda  \beta_l }{\sqrt{2}} 
\im \bigg( \ARpe1 \AsRpe0 - \ARpa1 \AsRpa0 -\{ R \leftrightarrow L \}  \big) \, ,\\ 
\mK_{20}^\prime &=&\frac{P_{\Lambda_b}\alpha_\Lambda  \beta_l}{\sqrt{2}} 
\im \bigg( \ARSPm \AsRpe1 -\ARSPp \AsRpa1 +\{ R \leftrightarrow L \}\nn\\
&+& \ALSPm \AsRpe1 + \ARSPm \AsLpe1 -\{ \| \leftrightarrow \perp \}  
\bigg) \, ,\\ 
\mK_{20}^{\prime\prime} &=&0  \, ,\\
\mK_{21} &=&  \frac{P_{\Lambda_b} }{\sqrt{2}} \im \bigg( 
\ARpa1 \AsRpa0 + \ARpe1 \AsRpe0 +\{ R \leftrightarrow L \}  \bigg)   \, ,\\ 
\mK_{21}^\prime &=& 0 \, ,\\ 
\mK_{21}^{\prime\prime} &=&- 2\sqrt{2} P_{\Lambda_b}
\im \bigg( 
\ARpa1 \AsRpa0 + \ARpe1 \AsRpe0 +\{ R \leftrightarrow L \}  \bigg)   \, ,\\
\mK_{22} &=&- \frac{P_{\Lambda_b} \beta_l }{\sqrt{2}} \im \bigg( 
\ARpa1 \AsRpe0 + \ARpe1 \AsRpa0 -\{ R \leftrightarrow L \}  \bigg)   \, ,\\ 
\mK_{22}^\prime &=& \frac{P_{\Lambda_b} \beta_l }{\sqrt{2}} \im  \bigg(  
\ARSPm \AsRpa1 + \ARSPp \AsRpe1 +\{ R \leftrightarrow L \}\nn\\
&+& \ALSPm \AsRpa1 + \ARSPm \AsLpa1 +\{ \| \leftrightarrow \perp \}    \bigg) \, ,\\ 
\mK_{22}^{\prime\prime} &=&0  \, ,\\
\mK_{23} &=& -\frac{P_{\Lambda_b}}{\sqrt{2}}  \re \bigg( 
\ARpa1 \AsRpe0 + \ARpe1 \AsRpa0 +
\{ R \leftrightarrow L \}  \bigg)   \, ,\\ 
\mK_{23}^\prime &=& 0 \, ,\\ 
\mK_{23}^{\prime\prime} &=& 2\sqrt{2} P_{\Lambda_b} \re \bigg( 
\ARpa1 \AsRpe0 + \ARpe1 \AsRpa0 +
\{ R \leftrightarrow L \}  \bigg)    \, ,\\
\mK_{24} &=&  \frac{P_{\Lambda_b} \beta_l }{\sqrt{2}}   \re \bigg(
\ARpa1 \AsRpa0 + \ARpe1  \AsRpe0 -  \{ R \leftrightarrow L \}  \bigg)   \, ,\\ 
\mK_{24}^\prime &=&  \frac{P_{\Lambda_b} \beta_l }{\sqrt{2}}   \re \bigg(  
\ARSPm \AsRpe1 + \ALSPp \AsLpa1 + \{ R \leftrightarrow L \} \nn\\
&+& \ALSPm \AsRpe1 + \ARSPm \AsLpe1 +
\{ \| \leftrightarrow \perp \}    \bigg) \, ,\\ 
\mK_{24}^{\prime\prime} &=& 0 \, ,\\
\mK_{25} &=& \frac{P_{\Lambda_b} \alpha_\Lambda}{\sqrt{2}} \im \bigg(  
\ARpa1 \AsRpe0 + \ARpe1 \AsRpa0 + \{ R \leftrightarrow L \}  \bigg)   \, ,\\ 
\mK_{25}^\prime &=&0 \, ,\\ 
\mK_{25}^{\prime\prime} &=& 2\sqrt{2}P_{\Lambda_b} \alpha_\Lambda
\im \bigg(  
\ARpa1 \AsRpe0 + \ARpe1 \AsRpa0 + \{ R \leftrightarrow L \}  \bigg) \, ,\\
\mK_{26} &=&\frac{ P_{\Lambda_b} \alpha_\Lambda \beta_l }{\sqrt{2}} \im  \bigg( \ARpa1 \AsRpa0 +\ARpe1 \AsRpe0 - \{ R \leftrightarrow L \}  \bigg)   \, ,\\ 
\mK_{26}^\prime &=& -\frac{ P_{\Lambda_b} \alpha_\Lambda \beta_l }{\sqrt{2}}
\bigg( \ARSPm \AsRpe1 + \ARSPp \AsRpa1 +\{ R \leftrightarrow L \}\nn\\
&+& \ALSPm \AsRpe1 + \ARSPm \AsLpe1 + \{ \| \leftrightarrow \perp \}    \bigg) \, ,\\ 
\mK_{26}^{\prime\prime} &=& 0  \, ,\\
\mK_{27} &=& -\frac{ P_{\Lambda_b} \alpha_\Lambda }{\sqrt{2}}
\re \bigg( \ARpa1 \AsRpa0 + \ARpe1 \AsRpe0 + \{ R \leftrightarrow L \}  \bigg)   \, ,\\ 
\mK_{27}^\prime &=& 0 \, ,\\ 
\mK_{27}^{\prime\prime} &=& 2\sqrt{2} P_{\Lambda_b} \alpha_\Lambda \re \bigg( \ARpa1 \AsRpa0 + \ARpe1 \AsRpe0 + \{ R \leftrightarrow L \}  \bigg)  \, ,\\
\mK_{28} &=&  \frac{P_{\Lambda_b}\alpha_\Lambda  \beta_l}{\sqrt{2}}  \re \bigg( \ARpa1 \AsRpe0 + \ARpe1 \AsRpa0 -\{ R \leftrightarrow L \} \bigg)   \, ,\\ 
\mK_{28}^\prime &=& \frac{P_{\Lambda_b}\alpha_\Lambda  \beta_l}{\sqrt{2}} \re \bigg( \ARSPm \AsRpa1 + \ARSPp \AsRpe1 +\{ R \leftrightarrow L \}\nn\\
&+&\ALSPm \AsRpa1 + \ARSPm \AsLpa1 +  \{ \| \leftrightarrow \perp \}    \bigg) \, ,\\ 
\mK_{28}^{\prime\prime} &=& 0  \, ,\\
\mK_{29} &=&  P_{\Lambda_b}\alpha_\Lambda   \im \bigg(
\ARSPp \AsRSPm +\{ R \leftrightarrow L \}  \bigg)   \, ,\\ 
\mK_{29}^\prime &=& -P_{\Lambda_b}\alpha_\Lambda   \im  \bigg( 
\ARSPm A_{\perp t}^* -\ARSPp A_{\| t}^* -\{ R \leftrightarrow L \}  \bigg) \, ,\\ 
\mK_{29}^{\prime\prime} &=&- 2P_{\Lambda_b}\alpha_\Lambda   \im  \bigg(   
\ARpa0 \AsRpe0 + \ARSPp \AsLSPm +\{ R \leftrightarrow L \} \nn\\
&+& \ARpa0 \AsLpe0 + \ARpe0 \AsLpa0 +  \{ \| \leftrightarrow \perp \} \nn\\
&-&A_{\perp t} A_{\|t}^*
\bigg)  \, ,\\
\mK_{30} &=&- P_{\Lambda_b}\alpha_\Lambda
\im \bigg( \ARpa0 \AsRpe0 -\ARSPp \AsRSPm  +\{ R \leftrightarrow L \}  \bigg)   \, ,\\ 
\mK_{30}^\prime &=&-P_{\Lambda_b}\alpha_\Lambda
\im \bigg(  
\ARSPm A_{\perp t}^* - \ARSPp A_{\| t}^* -\{ R \leftrightarrow L \} 
\bigg) \, ,\\ 
\mK_{30}^{\prime\prime} &=&  2 P_{\Lambda_b}\alpha_\Lambda \im \bigg(   
\ARpa0 \AsRpe0 - \ARSPp \AsRSPm +\{ R \leftrightarrow L \} \nn\\
&-&\ARpa0 \AsLpe0 + \ARpe0 \AsLpa0 + A_{\perp t} A_{\| t}^* \nn\\
&-&\ALSPp \AsRSPm - \{ \| \leftrightarrow \perp \} 
\bigg)  \, ,\\
\mK_{31} &=& - \frac{P_{\Lambda_b}}{2} \bigg( 
|\ARSPp|^2 - |\ARSPm|^2 + \{ R \leftrightarrow L \}  \bigg)   \, ,\\ 
\mK_{31}^\prime &=& - P_{\Lambda_b}\alpha_\Lambda \re \bigg(
\ARSPm A_{\| t}^* -\ARSPp A_{\perp t}^* - \{ R \leftrightarrow L \} \bigg) \, ,\\ 
\mK_{31}^{\prime\prime} &=& - P_{\Lambda_b}\alpha_\Lambda \bigg( 
|\ARpa0|^2 - |\ARpe0|^2 +\{ R \leftrightarrow L \}\nn\\
&+& |A_{\|t}|^2 - |A_{\perp t}|^2 \nn\\
&+& |\ARSPp |^2 -| \ARSPm |^2 +\{ R \leftrightarrow L \}\nn\\
&+&2 \re (\ALpa0 \AsRpa0 - \ALpe0 \AsRpe0 -\ARSPm \AsLSPm + \ARSPp \AsLSPp)  \bigg)  \, ,\\
\mK_{32} &=& - \frac{P_{\Lambda_b}\alpha_\Lambda}{2} \bigg(
|\ARpa0|^2  - |\ARpe0|^2 +  \{ R \leftrightarrow L \} \nn\\
&+& |\ARSPm| ^2 - |\ARSPp |^2+ \{ R \leftrightarrow L \}  \bigg)   \, ,\\ 
\mK_{32}^\prime &=& -P_{\Lambda_b}\alpha_\Lambda \re  \bigg( 
\ARSPm A_{\| t}^* - \ARSPp A_{\perp t}^* - \{ R \leftrightarrow L \}  \bigg) \, ,\\ 
\mK_{32}^{\prime\prime} &=& -P_{\Lambda_b}\alpha_\Lambda  \bigg(
|\ARpe0 |^2 - |\ARpa0 |^2 +  \{ R \leftrightarrow L \} + |A_{\| t} |^2 - |A_{\perp t}|^2\nn\\
&+& | \ARSPp |^2 - |\ARSPm |^2 +  \{ R \leftrightarrow L \} \nn\\
&+& 2\re (\ALpa0 \AsRpa0 -\ARSPm \AsLSPm - \{ \| \leftrightarrow \perp \} )
\bigg)  \, ,\\
\mK_{33} &=&-\frac{ P_{\Lambda_b}\alpha_\Lambda }{4} \bigg( 
|\ARpa1|^2 - |\ARpe1|^2 + \{ R \leftrightarrow L \}  \bigg)   \, ,\\ 
\mK_{33}^\prime &=& 0 \, ,\\ 
\mK_{33}^{\prime\prime} &=&  P_{\Lambda_b}\alpha_\Lambda \bigg( 
|\ARpa1|^2 - |\ARpe1|^2 + \{ R \leftrightarrow L \}  \bigg)   \, ,\\
\mK_{34} &=&-\frac{ P_{\Lambda_b}\alpha_\Lambda }{2} \im \bigg(
\ARpa1 \AsRpe1 + \{ R \leftrightarrow L \}  \bigg)   \, ,\\ 
\mK_{34}^\prime &=&0\, ,\\ 
\mK_{34}^{\prime\prime} &=& 2P_{\Lambda_b}\alpha_\Lambda
\im \bigg(
\ARpa1 \AsRpe1 + \{ R \leftrightarrow L \}  \bigg) 
\, ,\\
\mK_{35} &=& 0 \, ,\\ 
\mK_{35}^\prime &=& P_{\Lambda_b}\alpha_\Lambda  \beta_l 
\im \bigg(  \ARSPp \AsRpa0  -\ARSPm \AsRpe0 +\{ R \leftrightarrow L \} \nn\\
&+& \ALpe0 \AsRSPm + \ARpe0 \AsLSPm + \ARSPp \AsLpa0 + \ALSPp \AsRpa0  \bigg) \, ,\\ 
\mK_{35}^{\prime\prime} &=& 0  \, ,\\
\mK_{36} &=& 0  \, ,\\ 
\mK_{36}^\prime &=& -P_{\Lambda_b}\alpha_\Lambda  \beta_l  
\re \bigg( \ARSPm \AsRpa0 - \ARSPp \AsRpe0 +\{ R \leftrightarrow L \} \nn\\
&+& \ALSPm \AsRpa0 + \ARSPm \AsLpa0 -\{ \| \leftrightarrow \perp \}    \bigg) \, ,\\ 
\mK_{36}^{\prime\prime} &=& 0 
\end{eqnarray}

\end{document}